\documentclass[aps,10pt,prd,groupedaddress,nofootinbib,notitlepage,eqsecnum]{revtex4-1}
\usepackage[utf8]{inputenc}
\usepackage{hyperref}
\usepackage{xcolor}
\usepackage{graphicx}
\usepackage{amsmath,amssymb,amsbsy,amstext,amsthm}
\usepackage{mathtools}
\usepackage{amsfonts}
\newenvironment{equations}{\equation\aligned}{\endaligned\endequation}

\def\beq{\begin{equation}}
\def\eeq{\end{equation}}
\def\ba{\begin{equations}}
	\def\ea{\end{equations}}
\def\bc{\begin{center}}
	\def\ec{\end{center}}

\begin{document}
	\title{\boldmath Vacuum birefringence and the Schwinger effect in (3+1) de Sitter
		}
		\author{
		%\textsc
		{Mariona Banyeres$^{a}$}}
		%\email{{...}@{...}}

		\author{
		%\textsc
		{Guillem Dom\`enech$^{b}$}}
		\email{{domenech}@{thphys.uni-heidelberg.de}}

		\author{
		%\textsc
		{Jaume Garriga$^{a}$}}
		\email{{jaume.garriga}@{ub.edu}}
	
		\affiliation{
			$^{a}$\small{Departament de F\'\i sica Qu\`antica i Astrof\'\i sica, i Institut de Ci\`encies del Cosmos,\\
Universitat de Barcelona, Mart\'\i \ i Franqu\`es, 1, 08028, Barcelona, Spain}
			\\
			$^{b}$\small{Institut f\"ur Theoretische Physik, Ruprecht-Karls-Universit\"at Heidelberg \\ Philosophenweg 16, 69120 Heidelberg, Germany}
			}
		
	\date{\today}
	
	\begin{abstract} 
In de Sitter space, the current induced by an electric field in vacuum is known to feature certain peculiarities, such as infrared hyperconductivity for light bosons in weak electric fields. Moreover, negative conductivity has been claimed to occur for light bosons in moderate electric fields, and for fermions of any mass in electric fields below a certain threshold. Furthemore, in the limit of large mass and weak electric field, the current contains terms which are not exponentially suppressed, contrary to the semiclassical intuition. Here we explain these behaviors, showing that 
most of the reported negative conductivity is spurious. First, we show that the terms which are not exponentially suppressed follow precisely from the local Euler-Heisenberg Lagrangian (suitably generalized to curved space). Thus, such terms are unrelated to pair creation or to the transport of electric charge. Rather, they correspond to non-linearities of the electric field (responsible in particular for vacuum birefringence). The remaining contributions are exponentially suppressed and correspond to the creation of Schwinger pairs. Second, we argue that for light carriers the negative term in the regularized current does not correspond to a negative conductivity, but to the logarithmic running of the electric coupling constant, up to the high energy Hubble scale. We conclude that none of the above mentioned negative contributions can cause an instability such as the spontaneous growth of an electric field in de Sitter, at least within the weak coupling regime. Third, we provide a heuristic derivation of infrared hyperconductivity, which clarifies its possible role in magnetogenesis scenarios.
\end{abstract}

	\maketitle

\section{Introduction}

It is a good exercise to test our knowledge of quantum field theory in curved space-time, specially in cases where we can compute the results exactly. The pair creation of charged particles due to a constant external electric field in de Sitter space (dS) is one of such cases. Interestingly, in an expanding space-time, pairs can be created with screening or anti-screening orientation relative to the electric field. These two channels are usually refered to as ``downward" and ``upward" tunneling, and both can be relevant depending on the parameters \cite{Garriga:1994bm}. For example, in addition to the standard regime where the electric field plays the dominant role, pairs can also be produced gravitationally with a separation comparable to the dS radius. Furthermore, light fields in de Sitter tend to have an unusual infared (IR) behaviour, which leads to interesting phenomenology. 

The renormalized expectation value of the current induced by a constant electric field in dS 
takes a particularly simple form in 1+1 dimensions, where it can be expressed in terms of elementary functions \cite{Frob:2014zka}. 
Somewhat surprisingly, the exact result can also be obtained by adding the contributions of all created pairs along their semiclassical trajectories, considering both upward and downward tunneling. The agreement is of course expected in the semiclassical regime, but it actually extends to the full range of parameters.
In the regime $mH\lesssim eE \ll H^2$, where $e$ is the electric charge, $m$ is the mass of the charge carrier, and $H$ is the inverse dS radius, one finds the phenomenon of bosonic IR hyperconductivity, where the current is inversely proportional to the applied electric field $E$. On the other hand, in the strong electric field limit $\lambda\equiv eE/H^2 \gg 1$, the results of Schwinger effect in flat space are recovered.

The extension to $3$+$1$ dimensions \cite{Kobayashi:2014zza,Hayashinaka:2016qqn} revealed further peculiarities.\footnote{The case of $2$+$1$ dS was investigated in Refs.~\cite{Stahl:2015gaa,Bavarsad:2016cxh}.}  The results are similar to the 1+1 dimensional case in the strong field limit, and in the regime of IR bosonic hyper-conductivity. However, some differences remain poorly understood in the limit of heavy mass and weak field, and also for small masses and weak to moderate electric fields. The first issue is that in $3$+$1$ dS, the renormalized current due to very heavy charge carriers at weak electric field contains terms which decay only as an inverse power of the mass, while in the $1$+$1$ dS case one recovers the expected exponential suppression (from gravitational pair creation). This behavior was first encountered in Refs.~\cite{Kobayashi:2014zza,Hayashinaka:2016qqn}, by using adiabatic regularization, and then confirmed in Ref.~\cite{Hayashinaka:2016dnt} by using point splitting. Also, see Ref.~\cite{Bavarsad:2017oyv} for the inclusion of a magnetic field. Recently, Ref.~\cite{Hayashinaka:2018amz} proposed a ``maximal'' subtraction to remove the power-law terms, departing from the approach used in Refs.~\cite{Kobayashi:2014zza,Hayashinaka:2016qqn,Hayashinaka:2016dnt}. The implications for magnetogenesis of the ``maximal'' subtraction have been investaged in Ref.~\cite{Stahl:2018idd}.

Another unexpected feature is that in $3$+$1$ dS one finds negative values for the renormalized current in certain parameter ranges. In scalar QED this occurs only for light charge carriers ($m\lesssim 3\times10^{-3} H$), and for a window of moderate values of the electric field \cite{Kobayashi:2014zza}. In fermionic QED, the current is always negative below a certain threshold of the electric field, which depends on the mass of the fermion. The threshold is sizable 
(for all values of the mass), and for heavy fermions it grows linearly with the mass \cite{Hayashinaka:2016qqn}.

The Schwinger effect in dS is also interesting from the cosmological point of view. There are indications that magnetic fields are present in the inter-galactic medium with a strength $|B_{0,\mathrm{Mpc}}|\gtrsim 10^{-15} G$ and a correlation length of the order of Mpc \cite{Neronov:1900zz,Tavecchio:2010ja,Tavecchio:2010mk,Taylor:2011bn}. If magnetic fields were found in the voids as well, this would further indicate that they may have a primordial origin. A natural explanation would be inflationary magnetogenesis (for recent reviews see \cite{Kandus:2010nw,Durrer:2013pga,Subramanian:2015lua}). In such scenarios, an electromagnetic field is typically produced through a coupling to a rolling scalar field, which causes the effective electric charge to be time dependent.
%Primordial magnetic fields would leave imprints in the Cosmic Microwave Background (CMB) temperature and polarization anisotropies \cite{Puy:1998sv,Jedamzik:1999bm,Kosowsky:2004zh,Yadav:2012uz,Bonvin:2013tba}. Currently, the Planck data \cite{Ade:2015cva} yields an upper bound of $|B_{0,\mathrm{Mpc}}|<10^{-9}\mathrm{G}$. 
Unfortunately, inflationary magnetogenesis faces several problems. In the simplest models \cite{Turner:1987bw,Ratra:1991bn,Bekenstein:1982eu,Bamba:2006ga}, one encounters either a backreaction of the electric field or a strong coupling regime at very early times \cite{Demozzi:2009fu}, before reaching values of the magnetic field comparable to the current lower bound. For recent %developments
attempts to overcome these issues see, e.g., Refs.~\cite{Suyama:2012wh,Fujita:2012rb,Fujita:2013pgp,Fujita:2014sna,Ferreira:2013sqa,Ferreira:2014hma,Motta:2012rn,Giovannini:2013rza,Tasinato:2014fia,Green:2015fss,Fujita:2015iga,Domenech:2015zzi,Fujita:2016qab,Sharma:2017eps}.
On the other hand, 
%the presence of 
light charged scalar fields could also shut-off the process through the Schwinger effect\footnote{The Schwinger effect has been also studied for $SU(2)$ gauge fields during inflation \cite{Lozanov:2018kpk}. It turns out that due to the isotropy of the background value of the $SU(2)$ gauge fields the contribution from the Schwinger effect is negligible} \cite{Kobayashi:2014zza}.
%Significant effort has been put to explain possible magnetic fields with correlation lengths of the order of Mpc. %It turns out that in the simplest of circumstances one either encounters a strong coupling at early times or back-reaction of the electric field (see [...] for alternatives that try to circumvent the problems).
 %Although the model understudy might be an oversimplification of the dynamics that occur during inflation, it is a good crude estimatio
So far, the focus has been in estimating the effect of pair creation in specific inflationary scenarios, such as anisotropic inflation, where a persistent electric field is created by the rolling scalar field (see \cite{Watanabe:2009ct,Ito:2017bnn,Kitamoto:2018htg}). Also see Refs.~\cite{Stahl:2016geq,Sobol:2018djj} for a study of the backreaction on the inflationary dynamics. %On the other hand
Alternatively, one might imagine that a regime of negative conductivity may lead to the \textit{spontaneous electrification} of de Sitter, without the need of ad-hoc couplings. During inflation, this might generate a long range electric field of sizable magnitude, with possible implications for magnetogenesis. With this motivation, we will here re-examine the nature of the negative terms in the regularized vacuum current, and consider the possibility that they might lead to such an instability. 

The plan of the paper is the following. In Section \ref{sec:intro} we briefly review the Schwinger effect in 3+1 dS. For simplicity we focus on scalar QED, although most of our subsequent discussion carries over to the fermionic case in a straightforward way. In Section \ref{sec:current} we compute the induced current, regularized in the Pauli-Villars scheme, and discuss its behavior in different limits. Formally, the result is in agreement with earlier calculations which used adiabatic regularization or point splitting. In this sense, the peculiar negative terms which are found in the current are {not an artifact} of regularization. 

In Section \ref{sec:heavypairs} we study in detail the large mass and weak field limit of the current. We will see that the terms which are only suppressed by inverse powers of the mass can be explicitly derived from the so-called Euler-Heisenberg (EH) Lagrangian, generalized to curved space. This is the effective action after integrating out the heavy field, and it includes a series of higher dimensional local operators, which lead to non-linearities in the propagation of the electromagnetic field. In what follows we shall loosely refer to these as the vacuum birefringence terms, since some of them are responsible for a polarization dependent speed of propagation in the presence of a background electric field \cite{birefringence}. The QED quantum effective action in $1$+$1$ dS was studied in Ref.~\cite{Cai:2014qba}. However, such power-law behavior of the current is not present in that case.

In Section \ref{sec:semiclassical} we compute the semi-classical current due to the nucleated pairs. We find that it agrees well with the current of Section \ref{sec:current} in the limit of large mass and weak electric field, after the subtraction of the EH terms. Incidentally, such current is found to be negative. This strange behaviour is not necessarily forbidden in curved space-time. The current could be negative if, initially, the nucleated particles in the pairs move slower than the expansion rate, effectively providing a current that flows opposite to the electric field in the expanding coordinates.

In Section \ref{sec:spontaneous} we consider the negative contributions to the current which been identified in the infrared. These occur for a parameter range where the semi-classical approximation is not valid, and the effective action is non-local \cite{Barvinsky:1990up,Dunne:2004nc,Codello:2012kq}. In this regime, it would be hard to disentangle pair creation contributions from birefringence terms. Rather, as we shall see, the origin of infrared negative terms in the renormalized current is different. For very light charge carriers, the energy scale which is relevant for pair creation is not their rest mass, but the Hubble scale. We argue that the infrared negative contributions simply correspond to the running of the electric charge up to such physical high energy scale.

In Section \ref{sec:hyper}, we present a heuristic derivation of the phenomenon of IR hyperconductivity. This highlights its relation with the Higgsing of the Maxwell field by the fluctuating light charged scalar, and also clarifies the possible relevance of this phenomenon in scenarios of magnetogenesis.

 We conclude our work and extend the discussion in Section \ref{sec:conclusions}. Details of the calculations can be found in the appendices. We will follow the conventions for EM and QED from Refs.~\cite{Srednicki:2007qs,Subramanian:2015lua} {and work in Planck units ($\hbar=c=G=1$).}

%In this paper, we study in detail the cosmological creation of heavy pairs. 

\section{Quick review of Schwinger effect in 3+1 dimensions\label{sec:intro}}

Consider a complex scalar field $\phi$ with charge $e$ minimally coupled to a $U(1)$ gauge field $A_\mu$ in a de Sitter background. 
We will work in the flat slicing and in conformal coordinates in which the metric reads
\begin{align}\label{eq:dS}
g_{\mu\nu}=a^2\eta_{\mu\nu}\quad\mathrm{with}\quad a=\frac{-1}{H\tau}\qquad (-\infty<\tau<0)\,,
\end{align}
where $a$ is the scale factor, $H^{-1}$ is the de Sitter scale and $\tau$ is the conformal time. The physical time $t$ is defined by $dt=ad\tau$. The action for scalar QED is given by
\begin{align}\label{eq:sqed}
{\cal L}_{sQED}[\phi,A]=-\left(D_\mu\phi\right)^*D^\mu\phi-m^2\phi^*\phi-\frac{1}{4}F_{\mu\nu}F^{\mu\nu}\,,
\end{align}
where $D_\mu=\nabla_\mu-i e A_\mu$, $m$ is the mass and $F_{\mu\nu}=\nabla_\mu A_\nu - \nabla_\nu A_\mu$. Here, we will be primarily interested in the case of a constant external electric field $E$, say in the $z$ direction. We may then choose the gauge field to be
\begin{align}\label{eq:amu}
A_\mu=\frac{E}{H^2\tau}\delta_\mu^z\,,
\end{align}
so that $F_{\mu\nu}F^{\mu\nu}=-2E^2=\mathrm{constant}.$\footnote{In order to dynamically sustain such constant electric field in an expanding background we would need to consider that there is a source term feeding the electric field.
%with an effective negative current. 
This can be done by considering a coupling of the gauge field to a slowly rolling scalar field
%which dominates the background and 
\cite{Watanabe:2009ct}. The total action then is given by
\begin{align}
{\cal L}={\cal L}_{sQED}[\phi,A]+{\cal L}_{\mathrm{source}}[\varphi,A]
\qquad\mathrm{where}\qquad {\cal L}_{\mathrm{source}}[\varphi,A]=-\frac{1}{2}\nabla_\mu\varphi\nabla^\mu\varphi-V(\varphi)-\frac{f^2(\varphi)-1}{4}F_{\mu\nu}F^{\mu\nu}
\end{align}
with $f(\varphi)\propto a^{-2}$ (see Refs.~\cite{Watanabe:2009ct,Kitamoto:2018htg}). This type of couplings have been considered in anisotropic inflation \cite{Watanabe:2009ct} and in inflationary magnetogenesis (for recent reviews see \cite{Kandus:2010nw,Durrer:2013pga,Subramanian:2015lua}). As it has been pointed out in Ref.~\cite{Kitamoto:2018htg}, it is important to note that, in such a case, the electric coupling is changing rapidly and that one should take this variation into account when computing the Schwinger current in these scenarios.\label{footnote2}} % or in terms of the potential (assuming slow-roll) \cite{Kitamoto:2018htg,Watanabe:2009ct}
%\begin{align}
%f(\varphi)=\exp\left[\frac{2}{M_{pl}^2}\int d\varphi\frac{V}{\partial_\varphi V}\right]\,.
%\end{align}
%For simplicity, here we will consider that the coupling to the slowly rolling scalar field plays little effect in the Schwinger pair production. 
 For the time being, we will not be concerned with %such details, concerning 
the origin of the electric field. Rather, we shall study its effect on the vacuum of the the charged scalar.

\begin{figure}
\centering
\includegraphics[width=0.4\columnwidth]{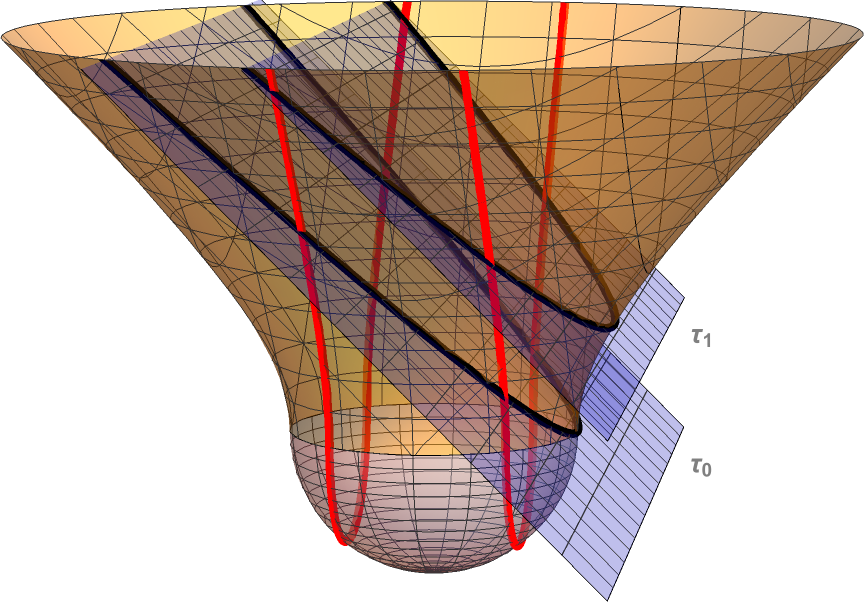}
\caption{Pair creation from electric fields in dS. The trajectories of the charged particles are shown in red. The lower (light red) part of the diagram is dS in the Euclidean manifold where the trajectories of charged particle are circular. The upper (light orange) part corresponds to the trajectories after analytic continuation to the Lorentzian manifold. The blue planes correspond to the flat slicing at a different times $\tau_0$ and $\tau_1$. We can understand $\tau_0$ as the time of nucleation in the flat slicing. At some later time $\tau_1$ the particles effectively moved in the direction of the electric field (black line in the flat slicing).\label{fig:0}}
\end{figure}

We proceed with the canonical quantization of $\phi$ in a constant electric field (see Appendix A for details on notation and conventions). After introducing the canonically normalized field, $q\equiv a\phi$, the equations of motion in Fourier space for the mode functions read \cite{Kobayashi:2014zza}
\begin{align}\label{eq:modefunctions}
q_k''+\omega_k^2q_k=0
\quad{,}\quad
\omega_k^2=k^2+2(aH)\lambda k r+(aH)^2\left(1/4-\mu^2\right)\,,
\end{align}
where $k^2=k_x^2+k_y^2+k_z^2$ is the co-moving wave-number, $q'\equiv \partial_\tau q$ and we introduced the notation \cite{Kobayashi:2014zza}
\begin{align}
\lambda=\frac{eE}{H^2}\quad,\quad r=\frac{k_z}{k}\quad{\mathrm{and}}\quad \mu^2=\frac{9}{4}-\frac{m^2}{H^2}-\lambda^2\,.
\end{align}
Note that $r<0$ is corresponds to ``downward'' tunneling and $r>0$ to ``upward'' tunneling. This notation comes from the semi-classical approximation where the tunneling rate is approximately given by the Euclidean action. In the Euclidean manifold charged particles follow circular trajectories that upon analytic continuation to the Lorentzian manifold become hyperbolic trajectories  (see Appendix~\ref{app:geodesics}). This is illustrated in Fig.~\ref{fig:0}.

The exact solution to Eq.~\eqref{eq:modefunctions} is given in terms of the Whittaker function \cite{Garriga:1994bm,Frob:2014zka,Kobayashi:2014zza} as
\begin{align}
q_k=\frac{\mathrm{e}^{-\pi\lambda r/2}}{\sqrt{2k}}W_{i\lambda r,\mu}(2ik\tau)\,.
\end{align}
It can be shown that $W_{i\lambda r,\mu}(2ik\tau)$ is the positive frequency solution in the asymptotic past ($-k\tau\to \infty$). 
%For later purposes, it is interesting to see that in the semi-classical limit where $\mu=i|\mu|$ and $|\mu|\gg 1$ (that is, the biggest contribution to the action comes form the classical equations of motion) 
It will be useful to decompose the mode function in terms of the Whittaker function $M_{i\lambda r,\mu}$ as
\begin{align}
q_k=\frac{\mathrm{e}^{i\pi\mu/2}}{2\sqrt{-ik\mu}}\left[\alpha_k M_{i\lambda r,\mu}(2ik\tau)+\beta_k \left(M_{i\lambda r,\mu}(2ik\tau)\right)^*\right]\,.
\end{align}
where $|\alpha_k|^2-|\beta_k|^2=1$. For imaginary values of $\mu$, which correspond to large mass or large electric field, the function $M_{i\lambda r,\mu}(2ik\tau)$ corresponds to the positive frequency solution 
with respect to cosmological time in the asymptotic future ($-k\tau\to 0$). 
In other words, in the limit where the semi-classical approximation is expected to hold, $\mu=i|\mu|$ and $|\mu|\gg 1$, there is also a well-defined adiabatic vacuum at future infinity. 
Thus $\alpha_k$ and $\beta_k$ are the Bogolioubov coefficients and in particular the number of created particles in the asymptotic future is given by \cite{Kobayashi:2014zza}
\begin{align}\label{eq:bogo}
|\beta_k|^2=\frac{\mathrm{e}^{-2\pi\lambda r}+\mathrm{e}^{-2\pi|\mu|}}{2\sinh\left(2\pi|\mu|\right)}\,.
\end{align}
In the semi-classical approximation one considers that the pairs are created at the time of maximum violation of the adiabatic condition.
%For given value of $k$, this occurs at around the time when \cite{Frob:2014zka,Kobayashi:2014zza} (%see Appendix \ref{app:adiabatic})
%\begin{align}\label{eq:break}
%k=k_c \sim aH|\mu|\,,
%\end{align}
For given value of $k$, this occurs at around the time when \cite{Frob:2014zka,Kobayashi:2014zza} %(see Appendix \ref{app:adiabatic})
\begin{align}\label{eq:cutoffmain}
{k}={k_c}\approx c \times aH |\mu|\,\qquad \mathrm{where}\qquad c\approx\left\{
\begin{array}{l}
1 \qquad \hspace{2.8mm}(\lambda \gg 1,m/H) \\
\, \\
\sqrt{2}\qquad (m/H\gg1,\lambda)
\end{array}\right.
\end{align}
and we introduced a constant $c$ in order to parametrize the uncertainty in the time of creation (see App.~\ref{app:adiabatic} for details). As we shall see, the numerical coefficient $c$ will be important in the semi-classical calculation since the result will depend on the cut-off. 
The total number of created particles per unit co-moving volume is then given by
\begin{align}
N=\frac{1}{(2\pi)^3}\int d^3 k |\beta_k|^2= \frac{1}{4\pi^2}\int_{-1}^1dr\int_0^{k_c} dk \,k^2 |\beta_k|^2 = \frac{c^3 a^3H^3|\mu|^3}{12\pi^2\sinh\left(2\pi\mu\right)}\left(\mathrm{e}^{-2\pi\mu}+\frac{\sinh\left(2\pi\lambda\right)}{2\pi\lambda}\right),
\end{align}
where we have used $k_c$ as the highest value of the momentum for which the pairs have been created at the given time. 
The total number of pairs created per unit physical volume is thus $n=N/a^3$. We will now proceed to compute the renormalized current and then we will compare it with the semi-classical approximation separately.

\section{Regularized expectation value of the induced current\label{sec:current}}
Functional differentiation of the action Eq.\eqref{eq:sqed} with respect to $A_\mu$ yields
\begin{align}\label{eq:eomgauge}
\nabla_\mu F^{\nu\mu}=J^\nu
\end{align}
where
\begin{align}
J_\mu = -\frac{ie}{2}\left\{\phi^\dag D_\mu\phi -\phi \left(D_\mu\phi\right)^\dag\right\}+\mathrm{h.c.}\,.
\end{align}
The expectation value of the current in the in vacuum state vanishes in all components except for the $z$-direction, which is given by
\begin{align}\label{eq:currentint}
\langle 0|J_z|0\rangle =\frac{2e}{a^2}\int \frac{d^3k}{(2\pi)^3}\left(k_z-eA_z\right)|q_k|^2\,.
\end{align}
By substituting the large argument expansion ($-k\tau\gg 1$) for the Whittaker function, i.e.
\begin{align}
W_{i\lambda r,\mu}(2ik\tau)=\mathrm{e}^{-2ik\tau}\left(2ik\tau\right)^{i\lambda r}\left(1+\frac{\mu^2-\left(\lambda+1/2\right)^2}{2ik\tau}+O((2ik\tau)^2)\right)\,,
\end{align}
into Eq.\eqref{eq:currentint}, it can be seen that the expectation value presents both quadratic and logarithmic divergences in the momentum, which we temporarily sidestep by introducing an ultraviolet regulator $\zeta$. In terms of $\zeta$, the integral \eqref{eq:currentint} can be solved analytically by using the Mellin-Barnes integral representation for the Whittaker function. This procedure was first used in 1$+$1 dS \cite{Frob:2014zka} and later in 3$+$1 dS \cite{Kobayashi:2014zza}. We obtain the same result in 3$+$1 as Ref.~\cite{Kobayashi:2014zza} (which includes a detailed derivation). The expectation value of the current is given by
\begin{align}
\langle J_z\rangle &=  aH\frac{e^2E}{4\pi^2}\lim_{\zeta\to\infty}\Bigg[\frac{2}{3}\left(\frac{\zeta}{aH}\right)^2+\frac{1}{3}\ln \frac{2\zeta}{aH}-\frac{25}{36}+\frac{\mu^2}{3}+\frac{\lambda^2}{15}
+F\left(\lambda,\mu,r\right)\Bigg]\,,
\end{align}
where for convenience we have defined
\begin{align}\label{eq:F}
F\left(\lambda,\mu,r\right)\equiv&\frac{45+4\pi^2\left(-2+3\lambda^2+2\mu^2\right)}{12\pi^3}\frac{\mu \cosh\left(2\pi\lambda\right)}{\lambda^2\sin\left(2\pi\mu\right)}-\frac{45+8\pi^2\left(-1+9\lambda^2+\mu^2\right)}{24\pi^4}\frac{\mu \sinh\left(2\pi\lambda\right)}{\lambda^3\sin\left(2\pi\mu\right)}\nonumber\\&
+\mathrm{Re}\Bigg\{\int_{-1}^1dr\frac{i}{16\sin\left(2\pi\mu\right)}\left(-1+4\mu^2+\left(7+12\lambda^2-12\mu^2\right)r^2-20\lambda^2r^4\right)\times\nonumber\\&
\left(\left(\mathrm{e}^{-2\pi r\lambda}+\mathrm{e}^{2\pi i\mu}\right)\psi\left(\frac{1}{2}+\mu-ir\lambda\right)-\left(\mathrm{e}^{-2\pi r\lambda}+\mathrm{e}^{-2\pi i\mu}\right)\psi\left(\frac{1}{2}-\mu-ir\lambda\right)\right)\Bigg\}\,.
\end{align}
As expected, the current contains a term which is quadratic in $\zeta$, as well as the logarithmic term. Instead of using an adiabatic subtraction scheme as in Ref.~\cite{Kobayashi:2014zza}, which is not manifestly gauge-invariant,
%by introducing a hard cut-off, 
we employ the Pauli-Villars regularization \cite{Frob:2014zka} (see Ref.~\cite{Hayashinaka:2016dnt} for the point-splitting regularization). In Pauli-Villars one introduces additional auxiliary fields %by hand
which may have the wrong kinetic sign (thus regarded as ghosts) in order to cancel the divergences. The mass of the auxiliary fields is then sent to infinity, rendering them completely non-dynamical. This simple mechanism works well at 1-loop level and manifestly preserves gauge invariance. In our current case, we need to introduce $3$ extra heavy fields that satisfy
\begin{align}
\sum_{i=0}^3 \mathrm{sign}(i)=0 \quad \mathrm{and}\quad \sum_{i=0}^3 \mathrm{sign}(i) m^2_i=0\,,
\end{align}
where $i=0$ is the original field $\phi$. Two of the extra fields will have the wrong sign and one of them will have the normal sign. We choose $\mathrm{sign}(i)=(-1)^i$ and $m_0=m$ (the original scalar), $m^2_2=4\Lambda^2-m^2$ and $m_{1}^2=m_{3}^2=2\Lambda^2$, where $\Lambda$ is a large scale which will be sent to infinity and plays the role of the regulator. Note that in the presence of the auxiliary fields, the momentum integrals are finite, and the momentum regulator $\zeta$ drops out. Indeed, in the large mass limit we have
\begin{align}\label{eq:Jzi}
\langle J_z\rangle_i&\approx  aH\frac{e^2E}{4\pi^2}\lim_{\zeta\to\infty}\Bigg[\frac{2}{3}\left(\frac{\zeta}{aH}\right)^2+\frac{1}{3}\ln \frac{2\zeta}{aH}-\frac{25}{36}+\frac{\mu^2}{3}+\frac{3\lambda^2}{15}-\frac{1}{6}\ln\frac{m_i^2}{H^2}+O\left(\frac{m_i^2}{H^2}\right)\Bigg].
\end{align}
The regularized current is the sum of all contributions, i.e.
\begin{align}
\langle J_z\rangle_{\mathrm{reg}}=\lim_{\Lambda\to\infty}\sum_{i=0}^3\mathrm{sign}(i)\langle J_z\rangle_i\,,
\end{align}
which after a short algebra reads
\begin{align}\label{eq:current1}
\langle J_z\rangle_{\mathrm{reg}} &=  aH\frac{e^2E}{4\pi^2}\lim_{\Lambda\to\infty}\Bigg[\frac{1}{6}\ln \frac{\Lambda^2}{m^2}-\frac{1}{6}\ln\left(1-\frac{m^2}{4\Lambda^2}\right)+\frac{1}{6}\ln \frac{m^2}{H^2}-\frac{2\lambda^2}{15}+F\left(\lambda,\mu,r\right)\Bigg]
\end{align}
This expression no longer contains the momentum regulator $\zeta$, but only the mass regulator $\Lambda$. Now, the second logarithm goes to zero when $\Lambda\to\infty$, so we can drop it. The remaining divergence can be reabsorbed into a renormalization of the charge. To see that, we look at the equations of motion for the gauge field Eq.~\eqref{eq:eomgauge}. We find that for a constant electric field the left hand side of Eq.~\eqref{eq:eomgauge} reads 
\begin{equation}
\nabla^\mu F_{\nu\mu}=-2(aH)E\delta_\nu^z. \label{later}
\end{equation}
It should be noted that this term is actually the so-called Hubble friction, responsible for the dilution of the electric field lines with the expansion of the universe (as mentioned above, some source is needed to keep the field constant, as for instance a coupling to a rolling scalar). Interestingly, the term \eqref{later} has the same linear dependence in the electric field and $H$ as the first term in \eqref{eq:current1}. Thus, we can absorb the latter in a counter term in the action of the form
\begin{align}
{\cal L}_{\mathrm{ct}}[A]=-\frac{Z_3}{4}F_{\mu\nu}F^{\mu\nu},
\end{align}
where $Z_3=-\frac{e^2}{48\pi^2}\ln \frac{\Lambda^2}{m^2}$. In other words, the charge is renormalized to
\begin{align}
e^2_{\Lambda}=\frac{e^2}{1-\frac{e^2}{48\pi^2}\ln \frac{\Lambda^2}{m^2}}\,. \label{running}
\end{align}
This renormalization of the electric charge was also observed in the point-splitting scheme \cite{Hayashinaka:2016dnt}.
% \textcolor{red}{Note that we could have used $\ln \frac{\Lambda^2}{2H^2}$ in the renormalization. However, it is more reasonable to assume that the regularization does not depend on $H$ which would be the IR cut-off of the theory.}
Here, we point out that 
\begin{equation}
\beta_e\equiv \frac{d e_\Lambda}{d\ln\Lambda} =\frac{e_\Lambda^3}{48\pi^2},
\end{equation}
is indeed the known beta function for scalar QED \cite{Srednicki:2007qs}. {It is important to realize that in renormalizing the charge we have assumed that the relevant scale of the system is the mass of the charge carrier $m$. That is, $e^2$ represents the effective coupling at the energy scale corresponding to the mass $m$. In the infrared limit where $m/H\ll1$ the remaining logarithmic term (the third one in Eq.~\eqref{eq:current1}) gives a large negative contribution that can be eliminated by running the effective couping to the Hubble scale which we shall call $e_{\mathrm{H}}$. This is related to $e$ by
\begin{align}
e^2_{\mathrm{H}}=\frac{e^2}{1-\frac{e^2}{48\pi^2}\ln \frac{H^2}{m^2}}\,.
\end{align}
We shall come back to this issue in Section \ref{sec:spontaneous}, when we discuss the possibility of infrared negative conductivity.
} 

We can now take the cut-off to infinity and the regularized current reads
\begin{align}\label{eq:regularized}
\langle J_z\rangle_{\mathrm{reg}} &=  aH\frac{e^2E}{4\pi^2}\Bigg[\frac{1}{6}\ln \frac{m^2}{H^2}-\frac{2\lambda^2}{15}+F\left(\lambda,\mu,r\right)\Bigg],
\end{align}
in agreement with Refs.~\cite{Kobayashi:2014zza,Hayashinaka:2016dnt}. 
%%%%%%%%%%%%
\begin{figure}
\centering
%\subcaptionbox{Hola}
	{\includegraphics[width=0.49\columnwidth]{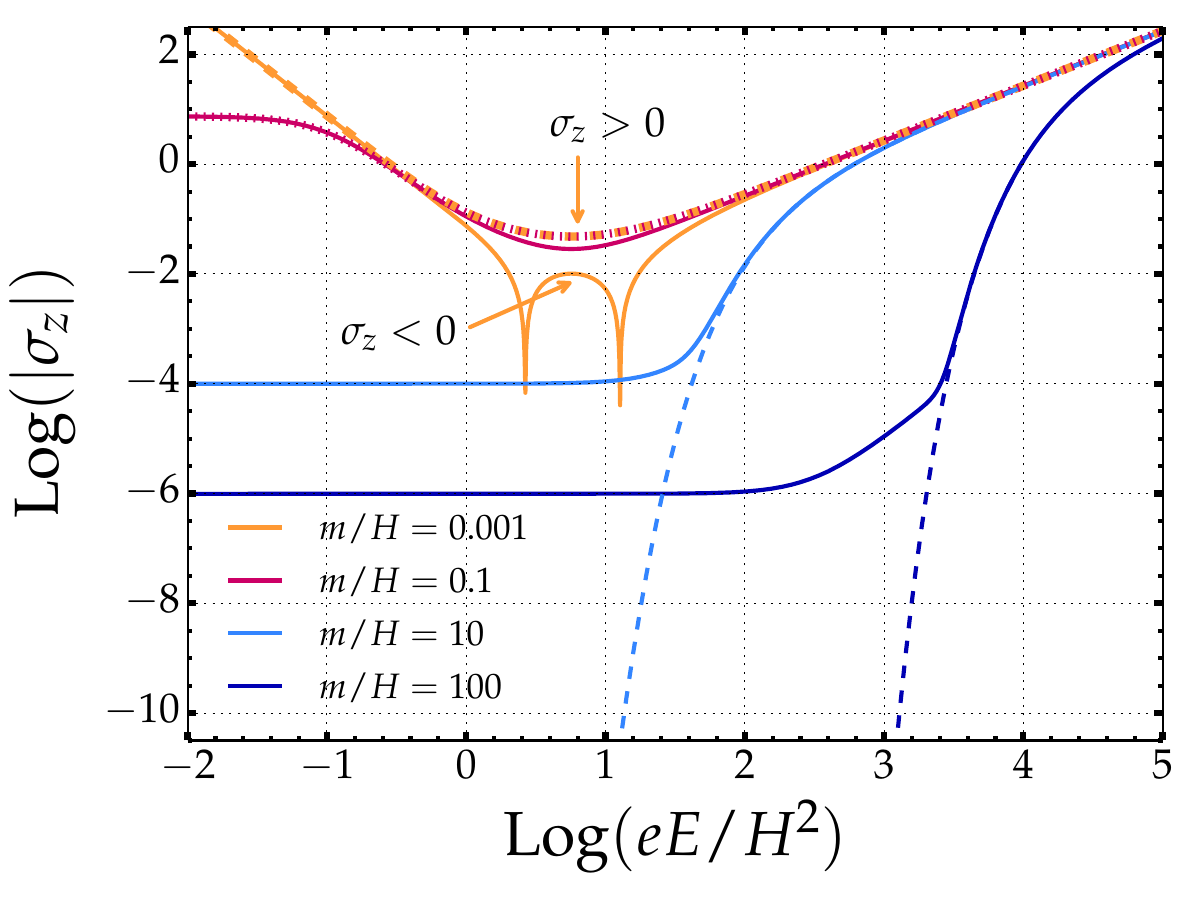}}
\hfill
\raisebox{1mm}{
%\subcaptionbox{Hola}
	{\includegraphics[width=0.49\columnwidth]{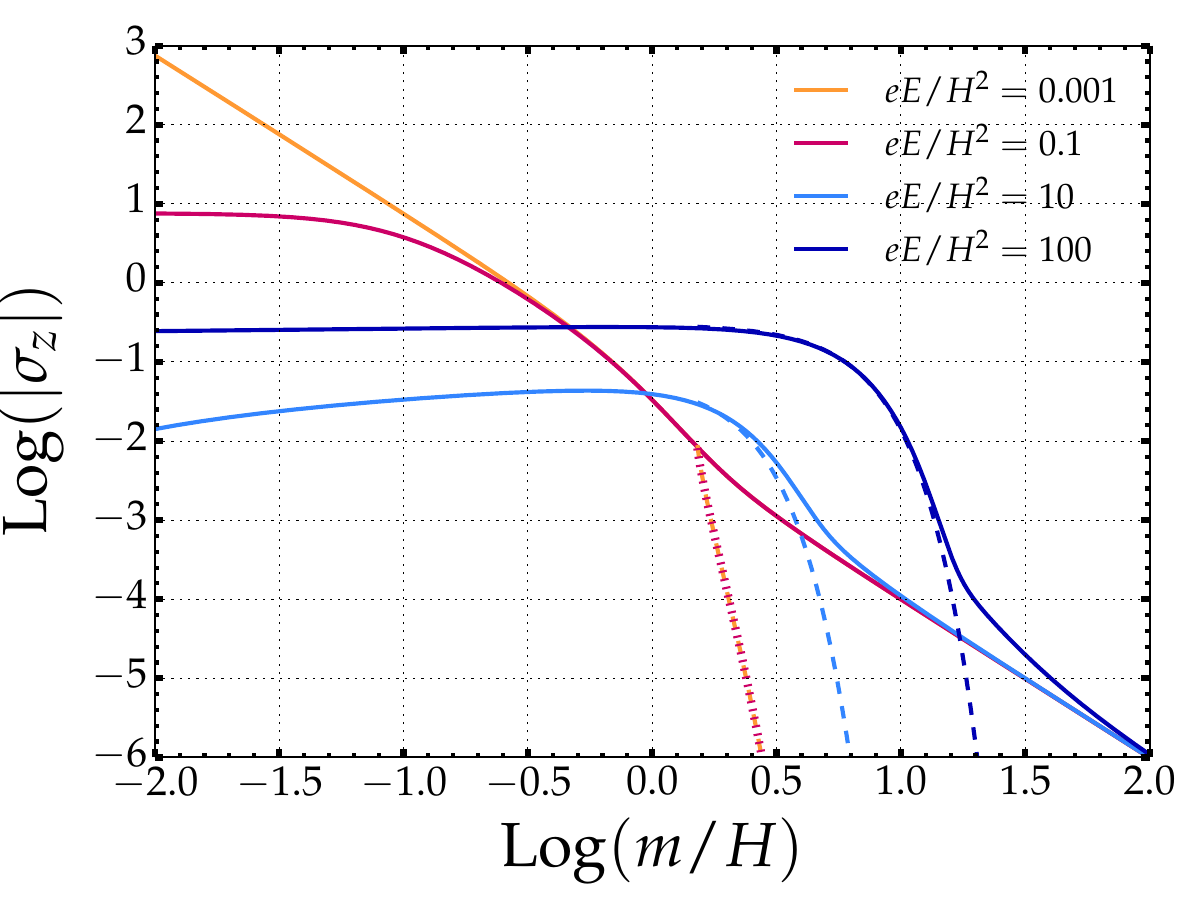}}
}
\caption{Absolute value of the dimensionless conductivity in Eq.~\eqref{eq:conductivity} for different parameter ranges. We plotted in full lines the exact formal result for the conductivity, from Eqs.~\eqref{eq:F} and \eqref{eq:regularized}. In dashed lines we show the expected exponential suppression for large masses and the conductivity after the renormalization of the charge for small masses. On the left hand side, the conductivity is plotted as a function of $\lambda=eE/H^2$ for different mass values. One can see that for $eE\gg H^2$ it approaches a linear behavior. Also note how for $m/H=0.001$ the conductivity becomes negative (between the two orange spikes) if we keep the logarithmic term in Eq.~\eqref{eq:regularized} (full orange line) but it is always positive if we do not (dotted orange line). On the right hand side, the conductivity is plotted as a function of the mass of the field. Note how for $m\gg H$ the values of the exact conductivity approach a power-law. \label{fig:1}}
\end{figure}
%%%%%%%%%%
We are ready to study the limits of the induced current. On one hand, in the strong field limit ($\lambda\gg 1,m/H$) one recovers the result from the Schwinger effect in flat space \cite{Kobayashi:2014zza}, i.e.
\begin{align}\label{eq:strongfield}
\langle J_z\rangle_{\mathrm{reg}}\approx\mathrm{sign}(E)\frac{aH}{12\pi^3}\frac{|e|^3E^2}{H^2}\mathrm{e}^{\frac{-\pi m^2}{|eE|}}\,.
\end{align}
Looking at Eq.~\eqref{eq:regularized}, this contribution comes from the $\cosh$ term in the second line. The real part of the integral in Eq.~\eqref{eq:F} (the second and third lines) gives a $2\lambda^2/15$ that cancels the second term of Eq.~\eqref{eq:regularized}. This result is in good agreement with the semi-classical approximation \cite{Frob:2014zka,Kobayashi:2014zza}, as we will see later in Section~\ref{sec:semiclassical}. 

On the other hand, if we expand around $m^2/H^2+\lambda^2\ll1$ we find \cite{Frob:2014zka,Kobayashi:2014zza}
\begin{align}
\langle J_z\rangle_{\mathrm{reg}}\approx aH\frac{3e^2E}{4\pi^2}\frac{1}{\frac{m^2}{H^2}+\lambda^2}\,. \label{hyper}
\end{align}
This limit of Eq.~\eqref{eq:regularized} (see also Eq.~\eqref{eq:F}) comes from the fractions with hyperbolic functions that contributes $5/6$ of the result and the real part of the integral contributes the remaining $1/6$. For $m^2/H^2\ll\lambda^2$ this corresponds to infrared hyperconductivity with $\langle J_z\rangle_{\mathrm{reg}}\propto E^{-1}$.

To illustrate the behavior of the current, we define the dimensionless conductivity
\begin{align}\label{eq:conductivity}
\sigma_z\equiv \frac{\langle J_z\rangle_{\mathrm{reg}}}{e^2E}\frac{1}{aH}.
\end{align}
Note that $J^\mu J_\mu = a^{-2} J_z^2$, so $J_z/a$ is actually the physical current. A numerical study shows that there is a narrow parameter region ($m/H< 3\times 10^{-3}$ and $\lambda\sim 1-10$) in which (\ref{eq:regularized}) becomes negative \cite{Kobayashi:2014zza,Hayashinaka:2016dnt}. Nonetheless, this is entirely due to the term proportional to $\ln(m^2/H^2)$ in (\ref{eq:regularized}), which as mentioned before can be absorbed in a redefinition of the coupling constant. Dropping this term, the bosonic current is positive for all values of parameters. In Fig.~\ref{fig:1} we plot the behavior of the conductivity for different parameters. In the left hand side, we can see how in the strong field limit all the conductivities come to the same (almost) linear behavior. Also for light enough fields (see the orange line with $m/H=0.001$) the conductivity becomes negative if we keep the logarithmic term (solid line), but it is everywhere positive if this term is dropped (dashed line). On the right hand side we can see that in the massive limit, the conductivities behave as a power of the mass. In both plots, the expected exponential behavior is show in dashed lines. In these limits where $|\mu|\gg1$, one expects the semi-classical approximation to work well but it seemingly does not \cite{Kobayashi:2014zza,Hayashinaka:2016dnt}. We will reconcile the two approaches in the following sections.

\section{Cosmological production of heavy pairs\label{sec:heavypairs}}

Let us now consider the limit of a large mass and weak to moderate electric field ($m/H\gg1$, $\lambda\lesssim 1$). In this case, one expects from the semi-classical intuition that the current would be exponentially suppressed as a function of the mass. Instead, one finds that the current contains a series of terms which are only suppressed by inverse powers of the mass \cite{Hayashinaka:2016dnt}. More precisely, we have
\begin{align}\label{eq:expansionJz}
\langle J_z\rangle_{\mathrm{reg}}\approx  aH\frac{e^2E}{4\pi^2}\Bigg[&\left(\frac{7}{18}-2{\xi}\right)\frac{H^2}{m^2}+\left(\frac{41}{90}+\frac{7}{180}\lambda^2\right)\frac{H^4}{m^4}+\left(\frac{676}{945}+\frac{19}{90}\lambda^2\right)\frac{H^6}{m^6}\nonumber\\&
+\left(\frac{401}{315}+\frac{2809}{3150}\lambda^2+\frac{31}{840}\lambda^4\right)\frac{H^8}{m^8}+\mathrm{O}\left(\frac{H^{10}}{m^{10}}\right)-\left(\frac{16\pi}{9}\frac{m^3}{H^3}+\mathrm{O}\left(\frac{m}{H}\right)\right)\mathrm{e}^{-2\pi \frac{m}{H}}\Bigg]\,,
\end{align}
where for later convenience, in this expression we allowed for a non-minimal coupling of the scalar field to gravity. In other words, in the standard result we replaced the mass $m$ by 
\begin{align}
m^2\to m^2+\xi R
\end{align}
and then we expanded the logarithmic term that appears in the function $F(\lambda,\mu,r)$ in the current \eqref{eq:current1} in the large mass limit (see Eq.~\eqref{eq:Jzi}), namely\footnote{We thank Lorenzo Ubaldi for pointing out that, in the previous version of the paper, the explanation accompanying this formula was misleading.}
\begin{align}
-\frac{1}{6}\ln \frac{m^2}{H^2}\to -\frac{1}{6}\ln \frac{m^2}{H^2}-\frac{1}{6}\frac{{\xi}R}{m^2}+\mathrm{O}\left(\frac{\xi^2R^2}{m^4}\right)\,.
\end{align}
The reason is that we will be interested in the lowest corrections to the renormalized current due to a non-minimal coupling to the curvature. 

\begin{figure}
\centering
\raisebox{0mm}{\includegraphics[width=0.25\columnwidth]{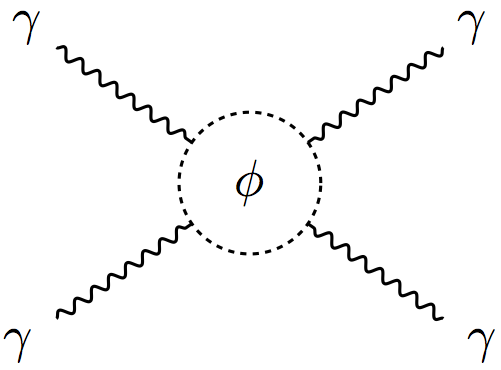}}
\hspace{2cm}
\raisebox{0mm}{
\includegraphics[width=0.25\columnwidth]{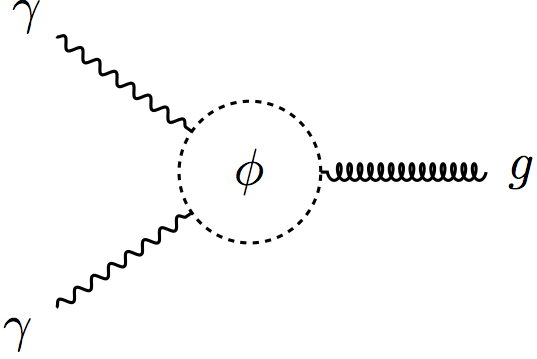}}
\caption{Examples of Feynman diagrams that contribute to the 1 scalar ($\phi$) loop quantum effective action. The left diagram is the 4 photon ($\gamma$) interaction and leads to vacuum birefringence. The right diagram is the 2 photon and 1 graviton ($g$) interaction. It does not lead to vacuum birefringence due to the symmetries of the background but subleading diagrams will.\label{fig:3}}
\end{figure}

First of all, note that within the square brackets in the left hand side of Eq.~\eqref{eq:expansionJz} there are terms that do not depend on the curvature $H$ at all. These are, for example, the terms with $\lambda^2 H^4/m^4=e^2E^2/m^4$ and $\lambda^4 H^8/m^8=e^4E^4/m^8$. This means that they will be present even in the flat space-time limit ($H\to 0$). Therefore, for $m^2\gg eE$, such terms should coincide with the effective non-linearities of the electric field after integrating out the massive field %In other words, we have to take into account the $n$-photon--$n$-photon (with $n>2$) interactions from integrating out one scalar loop 
(see Fig~\ref{fig:3}). These are encoded the effective action for scalar QED in flat space-time, which in the limit of large mass is the so-called Euler-Heisenberg Lagrangian (see Ref.~\cite{Dunne:2004nc} for a review), and its generalization to curved space. For the case of scalar QED this was given in Ref.~\cite{Bastianelli:2008cu}.
%%%%%%%%%%%%%%%%
%%%%%%%%%%%%%%%%%
The flat space Lagrangian and the leading correction due to spacetime curvature are respectively given by
\begin{align}\label{eq:1loopsca}
{\cal L}^{\mathrm{eff}}_{\mathrm{flat}}=\frac{e^4}{5760m^4\pi^2}\frac{1}{4}\left[7\Big(F_{\mu\nu}F^{\mu\nu}\Big)^2+\left(F_{\mu\nu}\tilde F^{\mu\nu}\right)^2\right]-\frac{e^6}{80640m^6\pi^2}F_{\mu\nu} F^{\mu\nu}\left[\frac{31}{4}\Big(F_{\mu\nu}F^{\mu\nu}\Big)^2+\frac{77}{16}\left(F_{\mu\nu}\tilde F^{\mu\nu}\right)^2\right]+...
\end{align}
and
\begin{align}\label{eq:1loopgra}
{\cal L}^{\mathrm{eff}}_{\mathrm{curv}}=\frac{e^2}{16\pi^2m^2}\left[\frac{1}{12}\left(\xi-\frac{1}{6}\right)RF_{\mu\nu}F^{\mu\nu}-\frac{1}{90}R_{\mu\nu}F^{\mu\alpha}F^\nu\,_\alpha-\frac{1}{180}R_{\mu\nu\alpha\beta} F^{\mu\nu}F^{\alpha\beta}+\frac{1}{60}\nabla^\alpha F_{\alpha\mu}\nabla_\beta F^{\beta\mu}\right]\,+...
\end{align}
where the ellipsis indicate higher dimension operators suppressed by higher powers of the mass. If we add these terms to the standard Maxwell Lagrangian, and in the presence of a classical current $J_{\mathrm{cl},\nu}$, we obtain the following equation of motion: 
\begin{equation}
\nabla^\mu F_{\mu\nu}-J^{\mathrm{eff}}_{\mathrm{EH},\nu}=J_{\mathrm{cl},\nu}\,,\label{cleom}
\end{equation}
where $J^{\mathrm{eff}}_{\mathrm{EH},\nu}\equiv J^{\mathrm{efff}}_{\mathrm{flat},\nu}+J^{\mathrm{eff}}_{\mathrm{curv},\nu}$. Here,
\begin{align}\label{eq:fscal}
J^{\mathrm{eff}}_{\mathrm{flat},z}=aH\frac{e^2E}{4\pi^2} \left[\frac{7}{180}\lambda^2\frac{H^4}{m^4}+\frac{31}{840}\lambda^4\frac{H^8}{m^8}\right]+...\ ,\quad
J^{\mathrm{eff}}_{\mathrm{curv},z}=aH\frac{e^2E}{4\pi^2} \left[\frac{7}{18}-2{\xi}\right]\frac{H^2}{m^2}\,+... 
\end{align}
Interestingly, the terms in Eq.~\eqref{eq:fscal} exactly match the terms number $1$, $2$, $4$ and $9$ from Eq.~\eqref{eq:expansionJz}. We expect to recover the remaining terms if we consider higher order corrections in the effective action. For example, operators of the form $(e^2/m^4)R^2 F_{\mu\nu}F^{\mu\nu}$ or $(e^4/m^6)R (F_{\mu\nu}F^{\mu\nu})^2$ would yield terms proportional to $(H^4/m^4)$ and $\lambda^2(H^6/m^6)$ respectively in the equations of motion for $A_\mu$. This leads us to conclude that all the terms which are an inverse power law of $m/H$ are due to the Euler-Heisenberg corrections from the quantum effective action. 

Although here we have focused on the case of scalar charges, a similar conclusion applies to fermions. In Appendix \ref{app:EHW} 
we confirm that the leading terms in the expansion in inverse powers of the mass which were found for fermionic currents in \cite{Hayashinaka:2016qqn}, can be derived from the EH lagrangian in fermionic QED (see Eqs.~\eqref{eq:1loopggraspin} and \eqref{eq:fspin}).

At this point, it should be noted that the correction terms ${\cal L}^{\mathrm{eff}}_{\mathrm{flat}}$ and ${\cal L}^{\mathrm{eff}}_{\mathrm{curv}}$ are gauge invariant, so $J^{\mathrm{eff}}_{\mathrm{EH},\nu}$ is identically conserved. For that reason, it cannot transport the actual charges which make up the current $J_{\mathrm{cl},\nu}$ in Eq. (\ref{cleom}), which will also have to be conserved by itself. 
As mentioned in the introduction, the EH terms generically lead to birefringence of the vacuum \cite{birefringence}. Note that in the quantum effective action we can write
\begin{align}
{\cal L}_{\mathrm{eff}}[A]=-\frac{1}{4}F_{\mu\nu}G^{\mu\nu\alpha\beta}F_{\alpha\beta}
\end{align}
where the constitutive tensor $G^{\mu\nu\alpha\beta}$ depends on the field strength as well as on the metric. In the presence of background electromagnetic fields, the constitutive tensor becomes nontrivial, and the propagation of linearized electromagnetic waves will depend on polarization. Technically, due to the maximal symmetry of de Sitter, the leading terms in  ${\cal L}^{\mathrm{eff}}_{\mathrm{curv}}$ cannot lead to birefringence, but subleading ones will.

For these two reasons, such contributions should not be attributed to pair creation, but should instead be interpreted as non-linearities in the kinetic term of the electromagnetic field. Thus, we conclude that the current generated through pair creation of heavy scalars is given by
\begin{align}\label{eq:paircurrent}
\langle J_z\rangle_{\mathrm{reg,pairs}}\approx -aHe^2E\frac{4}{9\pi}\frac{m^3}{H^3}\mathrm{e}^{-2\pi \frac{m}{H}}
\end{align}
%It should be noted that 
%Curiously, this expression has a negative sign, so heavy pairs actually contribute a current that flows opposite to the applied electric field. This is not so surprising in an expanding space-time. If the nucleated particles move initially slower than the expansion rate, the current will flow in the opposite direction of the electric field in the expanding coordinates.
Incidentally, this expression has a negative sign, so heavy pairs actually contribute a current that flows opposite to the applied electric field. This strange behavior is not necessarily forbidden in curved space-time. The current could be negative if, initially, the nucleated particles in the pairs move slower than the expansion rate, effectively providing a current that flows opposite to the electric field in the expanding coordinates.
%See again Fig.~\ref{fig:1}. 
The dashed lines in Fig.~\ref{fig:1} for $m>H$ represent the conductivity for the exact regularized current \eqref{eq:regularized} with all the power-law terms from Eq.~\eqref{eq:expansionJz} subtracted. This means that in practice in Fig.~\ref{fig:1} we have removed by hand the digamma function from Eq.~\eqref{eq:F} in Eq.~\eqref{eq:regularized} that leads to power-law terms. As we shall see, this is in qualitative agreement with the semi-classical approximation. 

Before closing this section, we would like to emphasize that the point of view here is very different from the maximal subtraction advocated in Ref.~\cite{Hayashinaka:2016dnt}, and so are the physical conclusions. In particular, we are not advocating to drop the contribution $J_{EH}$ altogether. Rather, this contribution is expected and should be maintained as a correction to the equations of motion for the electromagnetic field.

\section{Semi-classical current\label{sec:semiclassical}}

In this Section, we will consider the current due to pair creation, by adding the effect of all individual classical trajectories which are produced. The classical action is given by $S=-m\int ds+e\int A_\mu dx^\mu$, where $s$ is the invariant interval.
For a constant electric field in the $z$ direction, and in terms of the conformal time $\tau$, we have,
\begin{align}
S= \frac{m}{H}\int \frac{d\tau}{\tau} \sqrt{1-x'^2-y'^2-z'^2}+\lambda \int \frac{d\tau}{\tau}z',
\end{align}
from which the following first integrals are obtained:
\begin{align}\label{eq:geosol}
v^x= x'=\frac{-k_x\, \tau}{A}\quad,\quad v^y=y'=\frac{-k_y\, \tau}{A}\quad\mathrm{and}\quad v^z=z'=\frac{\lambda-k_z\,\tau}{A}.
\end{align}
Here, $v^i$ , with $i=x,y,z$, are the physical velocities relative to the Hubble flow, $k_i$ are integration constants which correspond to the conserved co-moving momenta and 
\begin{align}
A^2=\frac{m^2}{H^2}\gamma^2=\frac{m^2}{H^2}+\lambda^2+k^2 \tau^2-2k_z\lambda\tau\,, \label{A}
\end{align}
where we have used Eq.~\eqref{eq:geosol} in order to express the relativistic factor $\gamma = 1/ \sqrt{1-v^2}$ in terms of 
$\tau$. It is clear from \eqref{eq:geosol} that $k_z>0$ corresponds to ``upward'' tunneling and $k_z<0$ corresponds to ``downward'' tunneling, since only the latter has a turning point for $z$ at the value of conformal time given by $\tau=k_z/\lambda<0$. 
%With that we find that the 4-velocity of the particle is given by
%\begin{align}
%u^0=\frac{A}{a}\frac{H}{m}
%\end{align}
%so since $p^i=mu^0x'^{i}$ we have
The physical momenta are given by $p^i = m \gamma x'^i$, and we have
\begin{align}
p^x= \frac{k_x}{a},\quad p_y = \frac{k_y}{a}, \quad p^z= \frac{k_z+aH\lambda}{a}\,. \label{momenta}
\end{align}
For particles with $k_z<0$, $p_z$ is initially negative, becoming positive after the turning point. In terms of the physical momenta, the relativistic factor is given by $\gamma=(1+p^2/m^2)^{1/2}$.

Let us now compute the current $J_{\mu,\mathrm{sc}}$ due to semiclassical pair creation. It consists of two pieces, one from the trajectories of the pairs after they have been nucleated, and the other is a virtual current that links the two particles at the moment of nucleation. This additional piece is necessary for local charge conservation \cite{Frob:2014zka}. Hence, we write 
\begin{equation}
J_{\mu,\mathrm{sc}}=J_{\mu,\mathrm{pairs}}+J_{\mu,\mathrm{vir}}.
\end{equation} 
The first term can be computed by integrating the current that each pair generates, namely
\begin{align}\label{eq:paircurrent2}
J_{z,\mathrm{pairs}}=2ea\int v^z\, dn\approx \frac{e}{2\pi^2a^3}\int_{-1}^1dr\int_0^{k_c} dk \,k^2 |\beta_k|^2 \frac{\lambda-r k\tau}{\sqrt{|\mu|^2 +k^2\tau^2-2rk\lambda\tau}}.
\end{align}
Here $dn$ is the differential number density of pairs, that is $dn=(2\pi a)^{-3} d^3k |\beta_k|^2$, and $v^z\equiv p^z/m\gamma$ is the velocity of the nucleated particles in the $z$ direction given in Eq.~\eqref{eq:geosol}. Also, we approximated $|\mu|^2 \approx (m/H)^2+\lambda^2$ in the semi-classical regime, since $|\mu|\gg 1$. 
%We will take the cut-off to be $k_c\approx c|\mu| aH$ (see Appendix \ref{app:adiabatic}).
%Note that we have introduced the cut-off ($k_c\sim aH|\mu|$) when the adiabatic condition breaks down and the pairs are nucleated (see Eq.~\eqref{eq:break}). In the redefined variables $r\equiv k_z/k$ and $\upsilon\equiv-k\tau$ we have 
%\begin{align}\label{eq:pairs}
%J_{\mathrm{pairs},z}=aH\frac{eH^2}{2\pi^2}\int_{-1}^1dr\int_0^{|\mu|} d\upsilon \,\upsilon^2 |\beta_k|^2\frac{-\upsilon r+\lambda}{\sqrt{\upsilon^2-2\upsilon\lambda r+|\mu|^2+\frac{1}{4}}}\,.
%\end{align}
%In the large mass ($|\mu|\gg 1$) and weak field ($\lambda\ll 1$) limit we obtain
Expanding the integrand to first order in $\lambda\ll 1 $ and then performing the integrals we have
\begin{align}
J_{z,\mathrm{pairs}}\approx -(aH)e^2E\left\{1-\left(1-\frac{c^2}{2}\right)\sqrt{1+c^2}\right\}\frac{4}{9\pi}\frac{m^3}{H^3}\mathrm{e}^{-2\pi \frac{m}{H}}\,. \label{jpairs}
\end{align}
where we have used the value $k_c = c\,aH|\mu|$ for the momentum cut-off (see Eq.\eqref{eq:cutoffmain}), since pairs with higher momenta have not yet been created at time $\tau$. In Appendix \ref{app:adiabatic} we show that in the limit we are considering, the time of creation when the non-adiabaticity is the largest corresponds to $k_c \approx \sqrt{2} |\mu| a H$, and therefore $c=\sqrt{2}$. Substituting in Eq.~\eqref{jpairs} we find 
\begin{equation}
J_{z,\mathrm{pairs}} \approx -(aH)e^2E\frac{4}{9\pi}\frac{m^3}{H^3}\mathrm{e}^{-2\pi \frac{m}{H}}\,. \label{jpairsr}
\end{equation}
Surprisingly, this equation matches the field theory result \eqref{eq:paircurrent} to the dot, after subtraction of the Euler-Heisenberg terms. The reason we find this surprising is that we have not yet considered the contribution of the virtual current $J_{z,\mathrm{vir}}$ linking the pair at the time of creation. In the 1+1 dimensional case, a cancellation between both contributions was necessary in order to obtain the field theory result, but here we obtain it just from the contribution of the real pairs.

%%%%%%%%%%
\begin{figure}
\centering
\includegraphics[width=0.49\columnwidth]{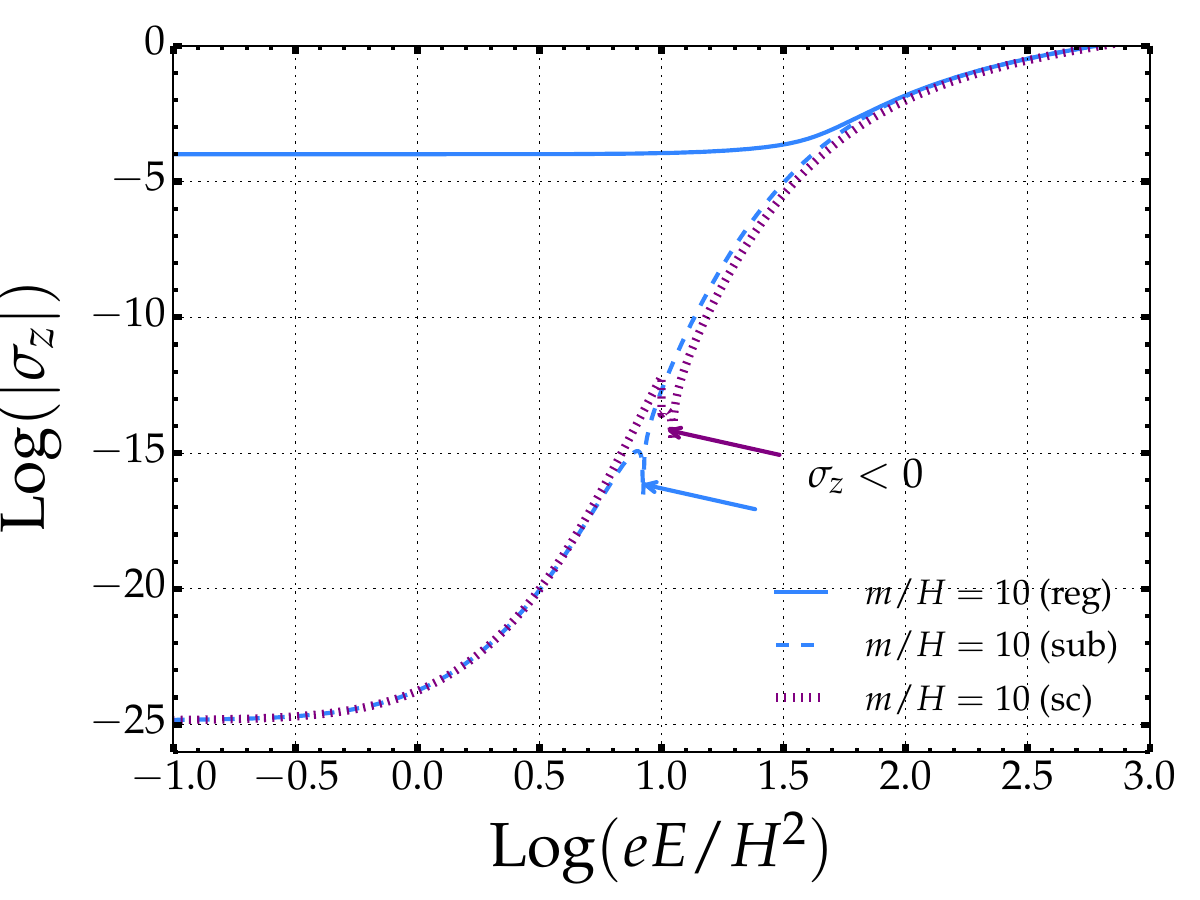}
\hfill
\raisebox{0mm}{
\includegraphics[width=0.49\columnwidth]{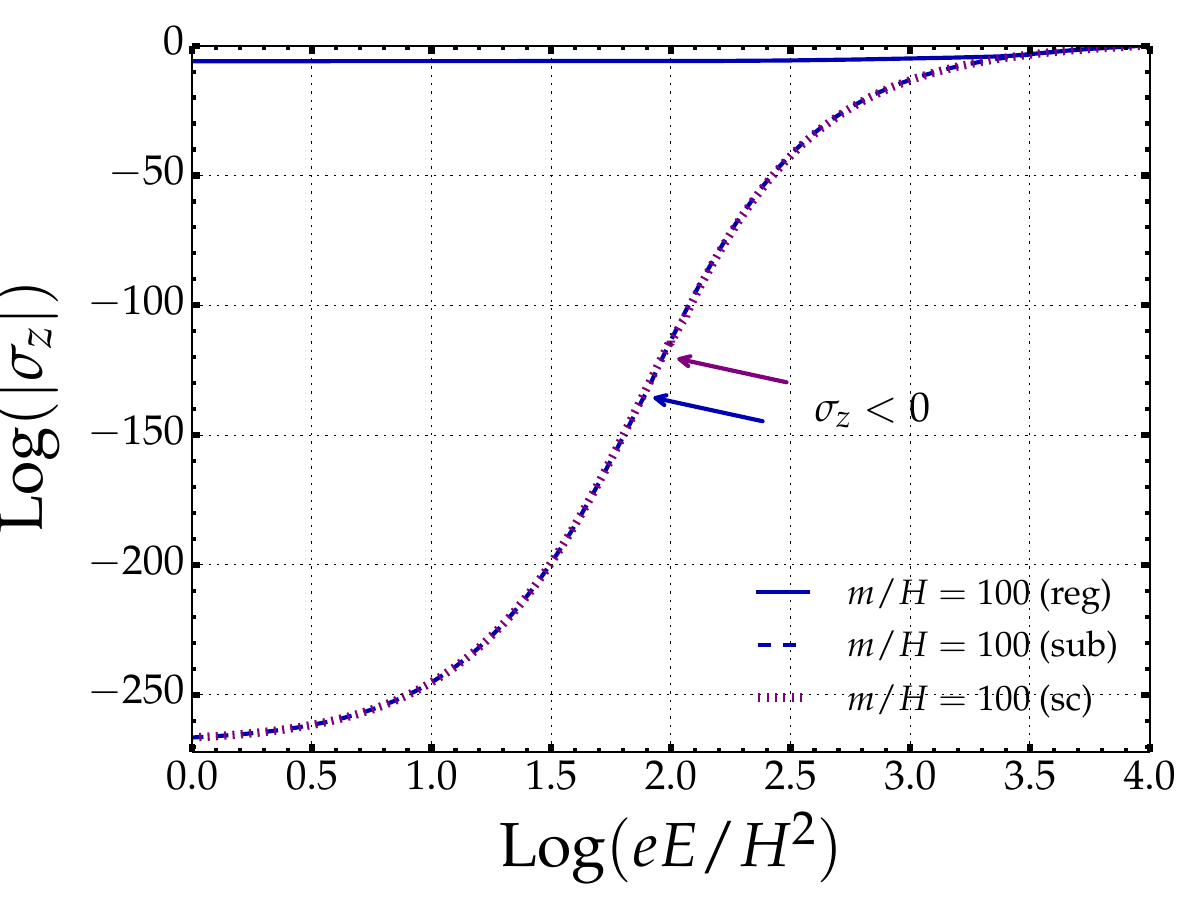}}
\caption{Absolute value of the conductivity as a function of $\lambda=eE/H^2$. Full lines correspond to the regularized conductivity (reg), dashed lines to the regularized conductivity with the digamma function subtracted (sub) and dotted lines correspond to the semi-classical approximation (sc). We indicated with an arrow the point where $\sigma_z$ changes sign ($\sigma_z>0$ at the right and $\sigma_z<0$ at the left  of the point). In the semiclassical approximation we have used only Eq.~\eqref{eq:paircurrent2} with $k_c=aH|\mu|$ for $\lambda>m/H$ and $k_c=\sqrt{2}aH|\mu|$ for $\lambda < m/H$. Because of this choice there is a change of sign at $\lambda=m/H$. See how the subtracted and semiclassical current (from the pairs only) agree quite well for all values of the electric field. On the left hand side we have $m/H=10$ and on the right hand side $m/H=100$.\label{fig:2}}
\end{figure}
%%%%%%%%%%%%%%%%%

For the contribution of the virtual current we consider a space-like world-line connecting the two charges at the moment of creation  \cite{Frob:2014zka}. For a pair nucleating at the time $t_p$ and centered at the point $x^i_p$, the virtual current is given by
\begin{align}
J_{\mathrm{vir},p}^\mu(x)=e\int ds \frac{dx^\mu}{ds}\frac{1}{a^3}\delta(t-t_p)\delta^{(3)}\left(x^i-x^i_p-\frac{a_0}{a} x_0^i(s)\right)
\end{align}
where $x^i_0(s)$ parametrizes the region where the current is non-vanishing, for a pair nucleating at the origin of coordinates at the time $t=0$. We take this region to be confined on the time slice $t=0$ and centered at $x^i=0$.  The constants $x_p^\nu$ correspond to spacetime translations of the same trajectory. The number of pairs which are created per unit physical time $t_p$ and unit co-moving volume, with $r=k_z/k$ in the interval $dr$ is given by
\begin{equation}
\frac{dN}{d^3 x_p dt_p}=\frac{dr}{(2\pi)^2} \frac{d}{dt_p} \int_0^{k_c} |\beta_k|^2 k^2 dk= \frac{H}{(2\pi)^2} dr |\beta_k|^2 k_c^3,
\end{equation} 
where in the last step we have used that $k_c \propto a(t_p)$, so $dk_c/dt_p = H k_c$. 
We can obtain the total current by adding the contribution of all pairs:
\begin{align}
\frac{dJ^i_{\mathrm{vir}}(x)}{dr}= \int \frac{dN}{dr} 
J_{\mathrm{vir,p}}^i(x)=  \frac{H}{4\pi^2}\frac{e}{a^3(t)} dr |\beta_k|^2  k_c^3(t) \int ds \frac{dx^i}{ds}
\end{align}
By symmetry, the only non-vanishing contribution will be in the $z$ direction, 
\begin{equation}
{J_{z,\mathrm{vir}}}= \frac{e a H}{4\pi^2} \int_{-1}^{1} dr |\beta_k|^2  d_{z,c} \left(\frac{k_c}{a}\right)^3.\label{nonva}
\end{equation}
Here,
\begin{align}
d_{z,c}=a(t)(\Delta z)_c
\end{align}
is the physical distance between the particles that nucleate at time $t$. For a given value of $r$, $d_{z,c}(r)$ is the same for all pairs at the time of their creation. 

In Appendix \ref{app:geodesics}, we argue that in the large mass $|\mu|\gg 1$ and weak field $\lambda\ll1$ limit which we are considering, the distance is given by
\begin{equation}
H  d_{z,c}\frac{k_c}{a} \approx -2r m\gamma_c -2H (1-r^2) f_z(r)   |\mu|\lambda [1+O(\lambda)]. \label{form}
\end{equation}
where $f_{z}(r)$ is a smooth even function, with $|f_z|\lesssim 1$. For $r=\pm 1$ the last term containing $f_z(r)$ is absent. In this case the motion is 1+1 dimensional, and the expression for the distance between particle and antiparticle at any given time can be found precisely from the analytic continuation of the Euclidean instanton. The same is true in the absence of electric field, $\lambda=0$. Unfortunately, for $\lambda\neq 0$ and generic values of $r$, there are no known instanton solutions. In fact, as we discuss in Appendix \ref{app:geodesics}, such solutions are unlikely to exist for $r^2\neq 1$, and hence we lack a method to determine $f(r)$ with precision. Nonetheless, we can still use the rough estimate \eqref{form} for the leading term in the large mass and weak field limit, based on the known limiting cases. 

Substituting \eqref{form}) into \eqref{nonva}, and after expansion in $\lambda$, the integration in $r$ is straightforward and we obtain
\begin{align}
J_{\mathrm{vir},z}\approx (aH)e^2E \left\{\frac{3}{2}\sqrt{1+c^2}-\frac{9\bar f}{8\pi}\right\}c^2\frac{4}{9\pi}\frac{m^3}{H^3}\mathrm{e}^{-2\pi \frac{m}{H}},
\end{align}
where, again, we have substituted $k_c\approx c aH |\mu|$.
Adding up both contributions we find that semi-classical approximation yields
\begin{align}\label{eq:totalsemi}
J_{\mathrm{sc},z}=J_{\mathrm{pair},z}+J_{\mathrm{vir},z}\approx-(aH)e^2E\left(1-(1+c^2)^{3/2}+\frac{9\bar f}{8\pi}c^2\right)\frac{4}{9\pi}\frac{m^3}{H^3}\mathrm{e}^{-2\pi \frac{m}{H}}\,,
\end{align}
Here we have introduced the constant $\bar f\equiv \int_{-1}^{1} (1-r^2)f(r)dr$, which also has the property $|\bar f|\lesssim 1$ and it does not depend on the cut-off, i.e. does not depend on $c$.
%Despite the imprecision in our estimate of $\bar f$, it should be noted that its value does not depend on the cut-off $c$. 
So we learn that, in contrast with the 1+1 dimensional case, here the total semiclassical current \eqref{eq:totalsemi} does depend on the cut-off. On the other hand, if we could by some method 
determine $\bar f$ more precisely, then we might expect to find for what value of $c$ does the semiclassical current agree with the field theory result, Eq. (\ref{eq:paircurrent}).  

Finally, let us recall that the criterion of maximum violation of adiabaticity leads to the value $c=\sqrt{2}$ for the cut-off. For that value  $J_{\mathrm{pair},z}$ given in (\ref{jpairsr}) already matches the field theory result for the exponentially suppressed part of the current, so we are strongly led to infer that if this value of $c$ is indeed a good estimate for the time of pair creation, then the virtual current (averaged over all orientations $r$ of the pair) is negligible. We should also stress that the method we have attempted here in order to evaluate this current worked very well in 1+1 dimensions. Nonetheless, in the present 3+1 dimensional context, generic pairs with $r\neq 1$ cannot be described by the analytic continuation of Euclidean solutions. For that reason, the semiclassical picture which we have adopted in order to calculate this piece may have missed some subtlety of the process of pair creation. Investigation of this issue is left for further research.

For the strong field limit ($\lambda\gg1,m/H$) we find that
\begin{align}
J_{\mathrm{sc},z}\approx\mathrm{sign}(E)\frac{aH\left(2-c^3\right)}{12\pi^3}\frac{|e|^3E^2}{H^2}\mathrm{e}^{\frac{-\pi m^2}{|eE|}}\,.
\end{align}
Using the value $c=1$ from Eq.~\eqref{eq:cutoffmain} (found in App.~\ref{app:adiabatic}) one recovers Eq.~\eqref{eq:strongfield}. In this regime, most of the pairs are nucleated with $r=- 1$ and at a short distance from each other compared with the Hubble scale, so the virtual current is negligible.

We plot in Fig.~\ref{fig:2} the comparison between the semi-classical current (in dotted lines) due to the pair production Eq.~\eqref{eq:paircurrent2} with the conductivity for the regularized current Eq.~\eqref{eq:regularized} (in full lines) and the regularized current with the Euler-Heisenberg terms \eqref{eq:expansionJz} subtracted (in dashed line). We plotted them only for $m/H>1$ as otherwise the semi-classical approximation breaks down and it is not clear how to identify the terms due to vacuum birefringence. In the calculation of the semiclassical current we have used that $k_c=aH|\mu|$ for $\lambda>m/H$ and $k_c=\sqrt{2}aH|\mu|$ for $\lambda<m/H$. It should be noted that have neglected the virtual currents for the reasons explained above. We approximated the result by matching at $\lambda=m/H$ and, thus, there is a change of sign at that value. Nevertheless, note that the change of sign is not so far from the change of sign of the EH subtracted current (dashed lines). We indicated with arrows the points where the conductivity flips sign and becomes negative. 

\section{Infrared negative conductivity?}\label{sec:spontaneous}

%In the previous sections, we have shown that the behaviour of the current in the limit of a large mass and a weak field can be well understood in the semi-classical limit. Also, we have computed the current using the Pauli-Vilars scheme, confirming earlier results. This indicates that there are no formal issues concerning the validity of the calculation. Nonetheless, i

In this section we would like to analyze whether negative infrared conductivity should really be taken seriously, or whether it may have any physical consequences. As pointed out in \cite{Hayashinaka:2016qqn}, in the presence of a negative conductivity the electric field may experience an instability. 
Here we would like to be more precise about this idea. 

Due to cosmic dilution, the instability will only be effective when the negative conductivity exceeds a certain threshold. Indeed, the equations of motion for the electric and magnetic fields measured by a cosmic observer, are given by 
 \cite{Subramanian:2015lua}
\begin{align}
\dot {E}^\kappa+2H{E}^\kappa=a^{-1}\,\mathrm{curl}(B)^\kappa-aJ^\kappa\qquad\mathrm{and}\qquad \dot {B}^\kappa+2H{B}^\kappa=-a^{-1}\mathrm{curl}(E)^\kappa\,,\label{friction}
\end{align}
where we have introduced the ``physical'' electric and magnetic fields  $E^\mu=a \bar E^\mu$ and $B^\mu=a\bar B^\mu$,
with $\bar E^\mu=F^{\mu\nu}u_\nu$, $\bar B_\mu=\frac{1}{2}\epsilon_{\mu\nu\alpha\beta}u^\nu F^{\alpha\beta}$ and
$u_\nu dx^\nu=-dt$. Also, $\mathrm{curl}(\bar B)^\kappa=\epsilon^{\kappa\beta\mu\nu}u_\nu\nabla_\beta \bar B_\mu$. 
Note in particular that the physical field which was used in previous sections is given by $E=E^z$, with upper index, while $F_{\mu\nu}F^{\mu\nu}=2(B^\mu B_\mu-E^\mu E_\mu)/a^2$. 

Let us for the moment consider the case of a homogeneous electric field, so we can also drop de magnetic term in (\ref{friction}).  Since $J^\kappa= a^{-1}\sigma e^2 H E^\kappa$, we find that the instability occurs when the current becomes negative with 
\begin{equation}
e^2 |\sigma| > 2 \label{threshold}
\end{equation}
Above that threshold, we would have an exponential growth of the electric field, at least in the linear regime.

Infrared logarithmic terms in the renormalized current may become negative, both for bosons and fermions.
The dimensionless conductivity of the induced current for fermions in the weak field ($\lambda\ll1$) and the small mass limit ($m\ll H$) is given by \cite{Hayashinaka:2016qqn}
\begin{align}
\sigma_\psi=\frac{1}{3\pi^2}\left(\ln \frac{m}{H}+\gamma_E-1/6+\mathrm{O}\left(\frac{m^4}{H^4}\right)\right)\,, \label{fermionconductivity}
\end{align}
where $\gamma_E \approx 0.577$ is Euler's constant. For example, if the instability were to take place during inflation, we can use $H\lesssim 10^{-6}M_{\mathrm{pl}} \sim2\times 10^{12} \mathrm{GeV}$, where $M^2_{\mathrm{pl}}=(8\pi G)^{-1}$, and light fermions with 
$m\gtrsim \mathrm{MeV}$, so that we have $m/H\gtrsim 10^{-15}$. Hence, the conductivity may easily be negative, but with 
$|\sigma_\psi| \lesssim 1$, which at weak coupling is well below the threshold (\ref{threshold}). 

Naively, one might think that in order to reach that threshold it is enough to increase the number of charged species. Their contributions would add up, and only a moderately large number of species would be needed,
\begin{equation}
N_\psi \gtrsim e^{-2} \sim 100.
\end{equation}
%charged fermions the conductivity is
%\begin{align}
%|\sigma|\approx \left(\frac{e}{10^{-1}}\right)^2 \frac{N_{\psi}}{3}\times 10^{-3}\left(0.41+\ln\frac{m_\psi}{H}+\mathrm{O}\left(\frac{m_\psi^4}{H^4}\right)\right)\,,
%\end{align}
However, in view of our discussion of the renormalization of the current in Section \ref{sec:current}, around Eq.\eqref{running}, it is clear that the coefficient in front of the logarithmic term is related to the renormalization of the electric charge. Indeed, we will now argue that in the regime where negative conductivity is supposed to occur, the theory has a ghost instability, even in flat space. 

First of all, we note that, repeating the same steps that we indicated in Section \ref{sec:current}, here we can absorb the logarithmic term into a running of the coupling constant, so that
\begin{equation}
e^2_\mathrm{H} = \frac{e^2}{1-\frac{N_\psi e^2}{12 \pi^2}\ln\frac{H^2}{m^2}}. \label{runningfermion}
\end{equation}
This running indeed corresponds to the standard beta function for fermionic QED in flat space $\beta^{\psi}_e = N_\psi e^3/(12\pi^2)$.
This is interesting, because the way we inferred in Section  \ref{sec:current} that the logarithmic terms renormalize the charge, is by noting that they have the same form as the friction term in Eq. (\ref{friction}), and hence can be reabsorbed in a wave function renormalizaton $Z_3$.

Hence, it should not come as a surprise that the condition (\ref{threshold}) is satisfied only above the energy corresponding to a Landau pole.  
We can indeed rewrite (\ref{runningfermion}) as
\begin{equation}
e^2_H = \frac{e^2}{1+e^2 \sigma_\psi/2}, \label{runningfermion2}
\end{equation}
where we use the expression \eqref{fermionconductivity} for infrared conductivity. For negative $\sigma_\psi$, and above the threshold \eqref{threshold} the denominator becomes negative, indicating that the kinetic term for the photon has flipped sign. We thus conclude that the instability of the electric field in de Sitter space is unrelated to any effects of curvature, but it is simply due to the fact that we are exploring the theory at energies which are above the Landau pole. In such regime, the theory is not well behaved, because the Maxwell field is a ghost. 

Ghosts are rather pathological in 3+1 dimensional local field theory, due to a catastrophic UV instability of the vacuum to production of pairs of positive energy particles accompanied by the production of ghosts . This pathology might be remedied by some drastic change in the UV structure of the theory, so that interactions become non-local above a certain energy scale. In this case, the vacuum might become long lived (see e.g. \cite{Garriga:2012pk} and references therein). For the sake of argument, let us side-step for the moment the UV catastrophe, and consider what would be the fate of a long wavelength ghost electric field.

Note that Eq.\eqref{fermionconductivity} is independent of the electric field for $\lambda\ll 1$. Taking the negative conductivity at face value, as in the analysis of Ref. \cite{Hayashinaka:2016qqn}, and for a standard kinetic term for the electric field, this would lead to the exponential instability
\begin{equation}
E \propto a^{|\sigma_\psi|-2}. 
\end{equation}
As a result, $\lambda$ would grow, the negative conductivity would decline, and a saturation value $E_*$ would be reached, with nearly vanishing conductivity. 
$|\sigma(E_*)|\approx 2$. The value of $E_*$ depends on the fermion masses, but in the mass range mentioned above we would have 
\cite{Hayashinaka:2016qqn} 
\begin{equation}
E_*\sim \frac{100}{e}H_*^2\,. 
\end{equation}
However, in view of our discussion above, this picture needs revision. Indeed, after we reabsorb the logarithmic term in Eq.\eqref{fermionconductivity} in a redefinition of the charge, the electric field becomes a ghost, and the remaining piece in the conductivity becomes positive
\begin{align}
\bar\sigma=\frac{N_\psi}{3\pi^2}\left(\gamma_E-1/6+\mathrm{O}\left(\frac{m^4}{H^4}\right)\right) >0\,. \label{fc2}
\end{align}
Since the electric field is now a ghost, this leads, for $\lambda\ll 1$, to the exponential instability
\begin{equation}
E \propto a^{\bar\sigma-2}\,, 
\end{equation}
provided that $\bar\sigma>2$. In Eq. (\ref{fc2}) we have only displayed the fermionic contributions, but in general the (positive) bosonic contributions should also be included. Further corrections which become important for $\lambda\gtrsim 1$ are also positive, and do not change the fact that $\bar\sigma>2$. In particular, for $\lambda\gg 1$ the conductivity grows linearly with $E$. Therefore, instead of reaching a saturation value $E_*$, the electric field grows superexponentially and without bound.

We conclude that the instability due to an apparent negative conductivity is actually a ghost instability, and that instead of leading to a saturation value for the electric field, it naturally leads to a catastrophic runaway growth. Of course, none of these catastrophes happen if we stay in the weak coupling regime along the RG flow from the scale $m$ to the scale $H$. In this case the Maxwell field is not a ghost and we do not have a spontaneous electrification of de Sitter.

\section{IR hyperconductivity and inflationary magnetogenesis.}\label{sec:hyper}

In this Section, we consider a heuristic derivation of Eq. (\ref{hyper}), which corresponds to the regime of IR hyperconductivity. As we shall see, this phenomenon is related to the Higgsing of the Maxwell field by the fluctuating charged scalar. Our discussion clarifies the possible relevance of hyperconductivity in scenarios of magnetogenesis. 

As it is well known, light fields in de Sitter tend to develop a large mean squared value. For a real scalar field in the Bunch Davies vacuum the expectation value is given by 
$\langle \phi^2 \rangle \approx 3H^4/(8\pi^2 m^2)$. If the scalar field is charged and the electric field is weak, the mass $m^2$ gets effectively replaced by $m^2\to m^2+e^2E^2/H^2$ \cite{Frob:2014zka} so that we have
\begin{equation}\label{eq:phivalue}
\langle \phi^*\phi \rangle \approx \frac{3}{4\pi^2} \frac{H^2}{\frac{m^2}{H^2}+\lambda^2}.
\end{equation}
The overall relative factor of $2$ is due to contributions from real and imaginary part of the charged field.
This expectation value gives a mass to the electromagnetic field given by
\begin{equation}\label{eq:Amass}
M_A^2 = e^2 \langle \phi^*\phi \rangle = \frac{3 e^2}{4\pi^2} \frac{H^2}{\frac{m^2}{H^2}+\lambda^2},
%\gg  H^2,
\end{equation}
which turns the medium into a superconductor. When the electric field is massive, it cannot exist in vacuum: it is quickly depleted through a current of the charged scalar field that gives mass to the $U(1)$ field. Indeed, we can derive Eq.~\eqref{hyper} by looking at the equation of motion for a massive gauge field, which reads 
\begin{equation}
\nabla_\mu F^{\nu\mu} = - M_A^2 A^{\nu}. 
\end{equation}
The mass term in this equation can be interpreted as an effective current $J_z = -M_A^2 A_z$. Then, for a constant electric field, with gauge potential $A_z = -a E/H$, this gives 
\begin{equation}
J_z = a (M_A^2/H) E, \label{jmass}
\end{equation}
from which Eq.~\eqref{hyper} follows after using \eqref{eq:Amass}. 

Let us now argue that hyperconductivity is not necessarily a major hindrance for magnetogenesis scenarios. The challenge in such scenarios is to provide a source for the electric field (such as for instance the coupling to a rolling scalar field described in footnote \ref{footnote2} of Section \ref{sec:intro}) which compensates for the cosmic dilution of the field lines, or ``Hubble friction''. This is represented by the term in the right hand side of Eq.~\eqref{later}. Comparing this to \eqref{jmass}, it is clear that the eff{}ect of hyperconductivity will be negligible relative to Hubble friction provided that
\begin{equation} 
M_A^2 \ll 2 H^2. \label{negligible}
\end{equation}
For instance, in the regime $(m/H)^2\ll \lambda^2 \ll1$ with $3e^2/(4\pi^2) \ll \lambda^2$, Eq.~\eqref{hyper} yields the hyperconducting behavior $J_z \propto E^{-1}$, but the mass of the gauge boson is very small $M_A^2 \ll H^2$, so the effect of the current is negligible.
On the other hand, for $3e^2/(4\pi^2) \gtrsim \lambda^2$ the electric field can be quite massive, $M_A^2\gtrsim H^2$. Naively, this seems to preclude the possibility of inflationary magnetogenesis. However, even in this second case, the actual dynamics can be more interesting. 

Indeed, note that
Eq. (\ref{eq:phivalue}) is just a stationary statistical average, which is only established after many e-foldings of inflation.
If the field starts near $\phi=0$ on the initial Hubble patch, then we have $\langle \phi^*\phi \rangle \approx H^3 t/(2\pi^2) \approx N H^2/(2\pi^2)$, where $N$ is the number of e-foldings since the beginning. The growth of the field is a Brownian process of step $\Delta \phi \sim  (H/2\pi)$ for each field component, which takes place each Hubble time and proceeds independently in different Hubble patches. The Brownian spreading is opposed by the classical drift due to the mass of the charged field, until the stationary distribution is established. Note, however, that if the number of e-foldings since the beginning of inflation is not too large, the effective mass of the gauge boson can still be very far from its stationary expectation value \eqref{eq:Amass}, and we have instead
\begin{equation}
M_A^2 \approx {e^2 N\over 2\pi^2} H^2.
\end{equation}
In this case, the effect of hyperconductivity will be negligible provided that
\begin{equation}
e^2 N \ll 2\pi^2,
\end{equation}
which is easy to achieve at weak coupling and with a moderate overall number of e-foldings.

We conclude that, as a matter of principle, a light charged scalar can cause the gauge boson to be very massive during inflation. This would preclude magnetogenesis in those regions where the charged scalar takes a large expectation value, due to its random brownian motion caused by quantum fluctuations. However, if the scalar field is near the origin at the time when the scale which corresponds to our visible universe first crossed the horizon, then the gauge field remains relatively light throughout inflation. In this case, the effect of hyperconductivity on magnetogenesis is completely negligible at weak coupling.

\section{Conclusions\label{sec:conclusions}}

We have reconsidered the Schwinger effect in de Sitter,
%While the exact calculation of the induced current in $1+1$ dimensions matched very well the semiclassical intuition \cite{Frob:2014zka}, in
addressing several puzzling features which seem to arise in different parameter ranges \cite{Kobayashi:2014zza,Hayashinaka:2016qqn,Hayashinaka:2016dnt}. In particular, we have explained the origin of the terms which are only power-law suppressed for large masses and weak fields (instead of having the expected exponential suppression). We have also clarified the nature of the 
negative currents which have been reported in the literature for light bosonic carriers in weak to moderate electric fields, and also for fermionic charge carriers of any mass below a certain threshold value for the electric field. In addition, we have provided a heuristic derivation of the regime of hyperconductivity, which is connected with the Higgsing of the gauge field by the fluctuating scalar.

We started by computing the induced current by a constant electric field in scalar QED, in Pauli-Villars regularization. Our results match those from the literature \cite{Kobayashi:2014zza,Hayashinaka:2016dnt}, confirming that the above mentioned peculiarities are not an artifact of regularization. 

We showed that for large masses and weak fields ($m/H\gg 1,\lambda$), the power-law suppressed terms in the induced current Eq.~\eqref{eq:expansionJz} correspond to the non-linearities of the electric field, which result from integrating out the massive charge carrier (see Fig~\ref{fig:2}). Indeed, we found that the leading order coefficients nicely match those coming from the Euler-Heisenberg Lagrangian, Eq.~\eqref{eq:fscal}, suitably generalized to curved space \cite{Dunne:2004nc,Bastianelli:2008cu}. This is true for scalar as well as fermionic carriers.  The latter are briefly discussed in Appendix \ref{app:EHW}, where we show that the leading non-linear contributions in the EH Lagrangian are negative (see Eqs.~\eqref{eq:1loopggraspin} and \eqref{eq:fspin}), and in agreement with the results of Ref.~\cite{Hayashinaka:2016qqn}.
The EH Lagrangian is an expansion of higher dimensional operators constructed from the electromagnetic and gravitational field, suppressed by the mass of the heavy charge carrier. 
Such operators are gauge invariant, and therefore their contributions to the current are identically conserved by themselves. In this sense, they do not represent the transport of actual electric charge from one place to another. Nonetheless, their contribution is expected and must be kept as a non-linear correction to the equations of motion for the electromagnetic field.  After these terms are subtracted, the remaining contributions correspond to the current induced by Schwinger pairs and are exponentially suppressed, as expected from semiclassical gravitational pair production \eqref{eq:paircurrent}.

To make this connection more quantitative, we computed the current in a semiclassical approximation. This consists of two pieces: the current carried  by Schwinger pairs in their classical trajectories, and a virtual current linking the pairs at the time of their creation. The latter is necessary for local charge conservation \cite{Frob:2014zka}. The results depend on the precise value of the time at which the pair is supposed to ``nucleate" in the semiclassical trajectory. This time is known quite precisely in the case of a strong electric field, where the bounce in the trajectory is well defined, but is less obvious in the case of cosmological pair creation. Following \cite{Frob:2014zka}, the moment of pair creation was determined by maximizing the non-adiabaticity parameter due to a time dependent frequency. In the regime ($m/H\gg 1,\lambda$), we find that the contribution of the classical trajectories matches quite precisely the field theory result, Eq.~\eqref{eq:expansionJz}, if we assume that the virtual current (averaged over all orientations of the pairs) is negligible. This result is somewhat puzzing, and it differs from the situation we encounter in 1+1 dimensions. There, the result turns out to be independent of the precise cut-off, which cancels out between the contribution from classical trajectories and virtual current. Unfortunately, a precise calculation of the virtual current in the present case is hindered by the fact that there are no instantons representing the nucleation of pairs with transverse momentum. Because of that, it is difficult to estimate with precision the distance between particle and antiparticle in a pair at the time of nucleation. This is left as a subject for further research.

%In general, we would need to know the exact value of the distance between the nucleated pairs. If that were the case, the matching between the exact results %and the semiclassical approach would give us the value of the cut-off of the nucleated particles. However, we argued that for pairs with $r\neq\pm1$ and $
%\lambda\neq0$, there is no closed Euclidean trajectories and thus we parametrized our ignorance of the virtual current with Eq.~\eqref{form}. On the other %hand, in the strong field limit ($\lambda\gg 1,m/H$) we also recovered the standard result, Eq.~\eqref{eq:strongfield}, using the cut-off of %Appendix~\ref{app:adiabatic}. We have shown in Fig.~\ref{fig:3} that in the large mass limit ($m/H\gg 1,\lambda$) the exact current \eqref{eq:regularized} %agrees well with the semiclassical calculation \eqref{eq:paircurrent} if we neglect the virtual current and we use the values for the cut-off \eqref{eq:cutoffmain}.

Let us now turn attention to the case of light charge carriers. We noted that for $m/H\ll1,\lambda$, the relevant energy scale is not the mass of the charge carrier $m$, but the de Sitter scale $H$. The logarithmic term in Eq.\eqref{eq:regularized} which is responsible for negative currents (for both scalar and fermions) \cite{Kobayashi:2014zza,Hayashinaka:2016qqn,Hayashinaka:2016dnt}, corresponds to the running of the electric coupling constant from the scale $m$ up to the high energy Hubble scale. In this sense, the presence of such terms is unrelated to any geometrical effect: the running proceeds with the standard flat space beta function. This prompted us to reexamine the phenomenon of negative conductivity  and whether it might lead to a growing instability. It was suggested in \cite{Kobayashi:2014zza,Hayashinaka:2016dnt} that an instability in the electric field might proceed up a saturation value $E_*(m)$ which depends on the mass of the charge carrier. By contrast, we argue that an instability would only happen if the negative logarithmic term is large enough to counteract the Hubble friction. However, we note that this would only happen if the inflationary Hubble scale were above the Landau pole, in which case the Maxwell field would behave as a ghost (see Eq.~\eqref{runningfermion2}). This would lead to an unbounded runaway instability the electric field, without saturation. We conclude that there is no spontaneous electrification of de Sitter, at least in the weak coupling regime where the Maxwell field has a positive kinetic term. 

%We have shown the behavior of the induced current in Fig.~\ref{fig:1}. 

Finally, we have provided a heuristic derivation of IR hyperconductivity, showing that this phenomenon is related to the Higgsing of the Maxwell field by the large mean squared value of the charged scalar during inflation. As a matter of principle, this causes the gauge boson to be very massive in regions where the scalar field value is large, precluding magnetogenesis from happening in those regions. However, if the scalar field is near the origin at the time when our visible patch first crossed the horizon, then (assuming weak coupling) the gauge boson remains relatively light throughout inflation. In this case the effect of hyperconductivity on magnetogenesis is quite negligible.

Throughout this paper we have considered the case where the electromagnetic coupling $e$ is constant. This is useful for comparison with existing calculations and also to investigate the possibility of spontaneous electrification. However, it is usually the case in magnetogenesis scenarios that the effective electromagnetic coupling changes rapidly in time. As mentioned in footnote \ref{footnote2} this case has been considered in Ref.~\cite{Kitamoto:2018htg} in the WKB approximation. This remains an interesting direction for future research.

\section*{Acknowledgments}
We would like to thank S.~Fl{\"o}rchinger, E.~Grossi and A.~Vilenkin for useful discussions and J.~Fedrow and J.~Takeda for comments on the draft. G.D. also thanks the MPA, Garching, for their hospitality while this paper was being written. This work was partially supported by DFG Collaborative Research center SFB 1225 (ISOQUANT)(G.D.) and FPA2016-76005- C2-2-P, MDM-2014-0369 of ICCUB (Unidad de Excelencia Maria de Maeztu), AGAUR 2014-SGR-1474, SGR-2017-754 (J.G.).

\appendix
\section{Quantization of the scalar field and mode functions\label{app:quantization}}

Here, we specify the notation and conventions which we use in the second quantization of the scalar field $\phi$. We promote the normalized fields $q$ and $q^*$ and their conjugate momenta into operators, namely
\begin{align}
q(\tau,\vec{x})=\frac{1}{(2\pi)^3}\int d^3k \left\{a_{\vec{k}}q_{\vec{k}}(\tau)\mathrm{e}^{i\vec{k}\cdot\vec{x}}+b^{\dag}_{\vec{k}}q^*_{-\vec{k}}(\tau)\mathrm{e}^{-i\vec{k}\cdot\vec{x}}\right\}\,,
\end{align}
and assign the commutation relations:
\begin{align}
\left[a_{\vec{k}},a^\dag_{\vec{p}}\right]=\left[b_{\vec{k}},b^\dag_{\vec{p}}\right]=(2\pi)^3\delta^{(3)}\left(\vec{k}-\vec{p}\right)\quad\mathrm{and}\quad
\left[a_{\vec{k}},a_{\vec{p}}\right]=[b_{\vec{k}},b^\dag_{\vec{p}}]=\left[a_{\vec{k}},b_{\vec{p}}\right]=[a_{\vec{k}},b^\dag_{\vec{p}}]=...=0\,.
\end{align}
To be compatible with the commutation relations of the field $q(\tau,\vec{x})$ and the canonical momenta $\Pi(\tau,\vec{x})$, the mode functions $q_k$ must satisfy the normalization condition
\begin{align}
q_k q^{'*}_k-q^*_k q^{'}_k=i\,.
\end{align}
The mode functions satisfy as well the equations of motion:
\begin{align}
q_k''+\omega_k^2q_k=0
\quad \mathrm{where}\quad
\omega_k^2=\left(k_z-eA_z\right)^2+k^2_x+k^2_y+a^2m^2-a''/a\,.
\end{align}
The vacuum in the asymptotic past is defined as $a_{\vec{k}}|0\rangle=b_{\vec{k}}|0\rangle=0$. In the asymptotic future (in the semi-classical approximation) we define the vacuum as $\hat{a}_{\vec{k}}|\hat 0\rangle=\hat{b}_{\vec{k}}|\hat 0\rangle=0$. They are related by
\begin{align}
\hat{a}_{\vec{k}}=\alpha_k{a}_{\vec{k}}+\beta^*_{k} \hat{b}^\dag_{-\vec{k}}\quad\mathrm{and}\quad\hat{b}_{\vec{k}}=\alpha_{-k}{b}_{\vec{k}}+\beta^*_{-k} \hat{a}^\dag_{-\vec{k}}
\end{align}
where \cite{Kobayashi:2014zza}
\begin{align}
\alpha_k=(-2i\mu)^{1/2}\mathrm{e}^{-(r\lambda+i\mu)\pi/2}\frac{\Gamma(-2\mu)}{\Gamma(1/2-\mu-ir\lambda)}
\quad \mathrm{and}\quad
\beta_k=-i(-2i\mu)^{1/2}\mathrm{e}^{-(r\lambda-i\mu)\pi/2}\frac{\Gamma(-2\mu)}{\Gamma(1/2+\mu-ir\lambda)}\,.
\end{align}
Therefore the total number of particles created in the asymptotic future is
\begin{align}
\langle \hat 0| \hat{a}^\dag_{\vec{k}}\hat{a}_{\vec{k}}|\hat 0 \rangle=\langle \hat 0| \hat{b}^\dag_{-\vec{k}}\hat{b}_{-\vec{k}}|\hat 0 \rangle=|\beta_k|^2\,.
\end{align}

\section{Breaking of the adiabatic condition\label{app:adiabatic}}
In this appendix we study the dependence of the cut-off with the parameters of the model. We start from the equations of motion, i.e.
\begin{align}
q_k''+\omega_k^2q_k=0
\quad \mathrm{where}\quad
\omega_k^2=\left(k_z-eA_z\right)^2+k^2_x+k^2_y+a^2m^2-a''/a\,.
\end{align}
It is convenient to work in physical time ($dt=d\tau/a$) and redefine the field as $\chi_k\equiv q_k a^{1/2}$. With these variables the previous equation becomes
\begin{align}
\ddot\chi_k+\Omega_k^2 \chi_k=0\quad \mathrm{where}\quad
\Omega^2_k=\omega_k^2a^{2}-H^2/4=\tilde m ^2 \left(1+\ell^2 (1+2r\upsilon+\upsilon^2)\right)
\end{align}
and we have introduced the notation
\begin{align}
\tilde m^2\equiv m^2-\frac{9}{4}H^2\quad,\quad \ell\equiv\frac{H\lambda}{\tilde m}\quad{\mathrm{and}}\quad\upsilon\equiv\frac{H}{\ell\tilde m}\left(-k\tau\right)\,.
\end{align}
Note that $\upsilon$ is positive definite. After a short algebra, the adiabatic condition reads
\begin{align}
f_k=\left|\frac{\dot\Omega_k}{\Omega_k^2}\right|=\frac{\ell^2H}{\tilde m}\frac{\left|\upsilon\left(r+\upsilon\right)\right|}{\left(1+\ell^2 (1+2r\upsilon+\upsilon^2)\right)^{3/2}}\,.
\end{align}
The function $f_k$ presents an extrema when
\begin{align}
\ell^2\left(\upsilon^2-1\right)=2-\frac{r}{\upsilon+r}+\ell^2\left(1-r^2\right)\frac{\upsilon}{\upsilon+r}\,.
\end{align}
The solutions in the limits $\ell\gg 1$ are given by
\begin{align}
\upsilon_+&\approx 1+\frac{3}{4\ell^2}+\mathrm{O}(\ell^{-3}) \qquad (r=1\,,\,\ell\gg1)\,,\\
\upsilon_-&\approx 1\pm\frac{1}{\sqrt{2}\ell} +\mathrm{O}(\ell^{-2}) \qquad (r=-1,\,\ell\gg1)\,,\\
\upsilon&\approx c_1+\frac{c_2}{\ell^2}+\mathrm{O}(\ell^{-3}) \qquad (r\neq \pm 1\,,\,\ell\gg1)\,.
\end{align}
where $c_1$ is a real positive solution to
\begin{align}
1-c_1^2+\frac{c_1}{c_1+r}\left(1-r^2\right)=0
\end{align}
and $c_2$ is related to $c_1$ by
\begin{align}
c_2=(c_1 + r)\frac{2 c_1 + r}{4 c_1 + r + 2 c_1^2 r + r^3}\,.
\end{align}
We can then calculate the cut-off to be
\begin{align}
\frac{k_c}{aH}&\approx\lambda+\frac{3}{4\lambda}\left(\frac{\tilde m}{H}\right)^2+\mathrm{O}(\lambda\ell^{-3})\qquad (r=1\,,\,\ell\gg1)\,,\\
\frac{k_c}{aH}&\approx \lambda\pm\frac{\tilde m}{\sqrt{2}H}+\mathrm{O}(\lambda\ell^{-2}) \qquad (r=-1,\,\ell\gg1)\,,\\
\frac{k_c}{aH}&\approx\lambda c_1+\frac{c_2}{2\lambda}\left(\frac{\tilde m}{H}\right)^2+\mathrm{O}(\lambda\ell^{-3})\qquad (r\neq\pm1\,,\,\ell\gg1)\,.
\end{align}
The solution in the limit $\ell\ll 1$ is given by
\begin{align}
\upsilon_\pm&\approx \sqrt{2}\ell^{-1}-\frac{r}{4}+\mathrm{O}(\ell)\qquad (\ell\ll1)\,.
\end{align}
There is another solution given by
\begin{align}
\upsilon_3&\approx-\frac{r}{2}-\frac{3r^3}{16}\ell^2+\mathrm{O}(\ell^3)\qquad (r<0\,,\,\ell\ll1)\,.
\end{align}
for which the adiabaticity value, say $f_k(\upsilon_3)$, is suppressed with respect to $f_k(\upsilon)$ by a power of $\ell$ and thus we can neglect it. The cut-off is then given by
\begin{align}
\frac{k_c}{aH}&\approx\sqrt{2}\frac{\tilde m}{H}-\frac{\lambda r}{4}+\mathrm{O}(\lambda\ell)\qquad (\ell\ll1)\,.
\end{align}
Thus, it is clear that the cut-off depends on $r$ and $\lambda$.
These approximations can be summarized as
\begin{align}\label{eq:cutoff}
\frac{k_c}{aH}\approx \left\{
\begin{array}{l}
|\mu| \qquad \hspace{4.5mm}(\ell \gg 1) \\
\, \\
\sqrt{2}|\mu|\qquad (\ell\ll 1)
\end{array}\right.\,.
\end{align}
In general we should write $k_c\approx c aH|\mu|$ where $c$ is an order one coefficient that depends on $\lambda$ and $r$. The coefficient is important in our case since the semi-classical calculation explicitly depends on the cut-off. For our purposes \eqref{eq:cutoff} will be enough though.
It should be noted that for $r=0$ there is an exact solution given by
\begin{align}
\upsilon_0=\sqrt{2}\sqrt{1+\ell^{-2}}\,.
\end{align}
In the $\ell\gg1$ limit the adiabatic value $f_k(\upsilon_0)$ is suppressed with respect to $f_k(\upsilon_\pm)$. This is expected since in the strong field regime most of the pairs will be nucleated with $r=\pm1$, i.e. oriented in the direction of the electric field. In the $\ell\ll 1$ limit $\upsilon_0$ coincides with $\upsilon_\pm$. This means that in absence of the electric field there is no privileged direction to be nucleated towards to.

\section{Equations of motion for Euler-Heisenberg\label{app:EHW}}
In this appendix we give the explicit expression for the terms that appear in the equations of motion due to the Euler-Heisenberg Lagrangians. The variation of Eqs.~\eqref{eq:1loopsca} and \eqref{eq:1loopgra} with respect to $A_\mu$ is respectively given by
\begin{align}
%\frac{1}{\sqrt{-g}}\nabla_\mu\left(\frac{\delta {\cal L}_{\mathrm{scal}}}{\delta \nabla_\mu A_\nu}\right)=
J^{\mathrm{eff},\nu}_{\mathrm{flat}}=\frac{7}{720}\frac{e^4}{m^2\pi^2}\nabla_\mu\left[F_{\alpha\beta}F^{\alpha\beta}F^{\mu\nu}\right]-\frac{31}{26880}\frac{e^4}{m^4\pi^2}\nabla_\mu\left[\left(F_{\alpha\beta}F^{\alpha\beta}\right)^2F^{\mu\nu}\right]
\end{align}
and
\begin{align}
%\frac{1}{\sqrt{-g}}\nabla_\mu\left(\frac{\delta {\cal L}_{\mathrm{scal}}}{\delta \nabla_\mu A_\nu}\right)=
J^{\mathrm{eff},\nu}_{\mathrm{curv}}=\frac{e^2}{16\pi^2m^2}\Bigg[\frac{1}{3}\left(\frac{1}{6}-\xi\right) \nabla_\mu\left(RF^{\mu\nu}\right)+\frac{1}{45}\nabla_\mu&\left(R^\mu\,_\alpha F^{\alpha\nu}+R^\nu\,_\alpha F^{\mu\alpha}\right)+\frac{1}{45}\nabla_\mu\left(R^{\mu\nu\alpha\beta}F_{\alpha\beta}\right)\nonumber\\&
-\frac{1}{30}\left(R^\nu\,_\alpha \nabla_\beta F^{\beta\alpha}-\nabla_\alpha\nabla^\alpha\nabla_\beta F^{\beta\nu}\right)\Bigg]
\end{align}
Using that 
\begin{align}
R=12H^2\quad\,\quad R_{\mu\nu}=3(aH)^2g_{\mu\nu}\quad{\rm and}\quad  R_{\mu\nu\alpha\beta}=2a^2(aH)^2g_{\mu[\alpha}g_{\beta]\nu}
\end{align}
we obtain Eq.~\eqref{eq:fscal} in the main text.

We also provide the leading correction from the curvature to the current in fermion QED. The quantum effective Lagrangian is given by \cite{Bastianelli:2008cu}
\begin{align}\label{eq:1loopggraspin}
{\cal L}^{\mathrm{eff}}_{\psi,\mathrm{curv}}=\frac{e^2}{16\pi^2m^2}\left[\frac{1}{36}RF_{\mu\nu}F^{\mu\nu}-\frac{13}{90}R_{\mu\nu}F^{\mu\alpha}F^\nu\,_\alpha+\frac{1}{90}R_{\mu\nu\alpha\beta} F^{\mu\nu}F^{\alpha\beta}+\frac{2}{15}\nabla^\alpha F_{\alpha\mu}\nabla_\beta F^{\beta\mu}\right]\,+...
\end{align}
and the induced current reads
\begin{align}
%\frac{1}{\sqrt{-g}}\nabla_\mu\left(\frac{\delta {\cal L}_{\mathrm{scal}}}{\delta \nabla_\mu A_\nu}\right)=
J^{\mathrm{eff},\nu}_{\psi,\mathrm{curv}}=\frac{e^2}{16\pi^2m^2}\Bigg[-\frac{1}{9} \nabla_\mu\left(RF^{\mu\nu}\right)+\frac{13}{45}\nabla_\mu&\left(R^\mu\,_\alpha F^{\alpha\nu}+R^\nu\,_\alpha F^{\mu\alpha}\right)-\frac{2}{45}\nabla_\mu\left(R^{\mu\nu\alpha\beta}F_{\alpha\beta}\right)\nonumber\\&
-\frac{4}{15}\left(R^\nu\,_\alpha \nabla_\beta F^{\beta\alpha}-\nabla_\alpha\nabla^\alpha\nabla_\beta F^{\beta\nu}\right)\Bigg]\,.
\end{align}
After a short algebra we find that
\begin{align}\label{eq:fspin}
J^{\mathrm{eff}}_{\psi,\mathrm{curv},z}=-aH\frac{e^2E}{36\pi^2}\frac{H^2}{m^2}\,+... \,,
\end{align}
which is exactly the leading negative behavior found in Ref.~\cite{Hayashinaka:2016qqn}.

\section{Classical trajectories for the created pairs\label{app:geodesics}}

In this Appendix, we discuss the solution of the equations of motion for the charged particle in a constant electric field, and we consider the 
the trajectories of charged pairs.

Our starting point are the first integrals of motion, Eqs.~\eqref{eq:geosol}, in terms of the conserved co-moving momenta $k_i$.
Due to symmetry of the electric field around the $z$ axis, and without loss of generality, we can always rotate in the transverse $(x,y)$ plane so that $k_y=0$. Then the trajectory will be on a $y=y_0=const.$ plane. Also, by rotation of $\pi$ we can change the sign of $k_x$, so in what follows we shall take $k_x>0$. In such coordinate system, integration of Eq.~\ref{eq:geosol} gives the solution
\begin{align}
k_x(x-x_0)&=-(1-r^2) \left\{A+r\lambda \ln\left[A+k \tau-r\lambda\right]- B_x\right\} \label{pair1} \\
k_z(z-z_0)&= -r^2 A+(1-r^2)  \left\{r\lambda \ln\left[A+k\tau-r\lambda\right]-B_z\right\}, \label{pair2}
\end{align}
where we used $r=k_z/k$, $k_x/k=1-r^2$, and $A=m\gamma/H$ is given in Eq.~\eqref{A}: 
\begin{equation}
A=\left(\frac{m^2}{H^2}+\lambda^2+k^2 \tau^2-2k_z\lambda\tau\,\right)^{1/2}.\label{A2}
\end{equation}
For $r=-1$ this solution agrees with the results of Refs.~\cite{Frob:2014zka,Bicak:2005yt}.
%where they used an interesting method to map constant velocity trajectories in Minkowski space-time to constant acceleration trajectories in dS.

The constants $B_i$ are actually redundant at this stage, as one could place them inside of $x_0$ or $z_0$ in the left hand side of the equations. 
The reason to introduce new constants here is because we want $(x_0, z_0)$ to represent the center of mass of a given particle-antiparticle pair.  In such case, $B_i$ would be determined if we knew what is the relative separation between particle and antiparticle, for given values of the particle's physical momentum $\vec p$. 
To illustrate this point, let us note that with the help of Eqs.~\eqref{momenta} the trajectory can be rewritten as 
\begin{align}
p^x d_x &=-2 (1-r^2)\left(\frac{m\gamma}{H}+r\lambda\left\{\ln\left[\frac{m\gamma r-p^z+(1-r^2) H\lambda}{m\gamma_c r-p_c^z+(1-r^2) H\lambda}\right]-\bar B_x(r,|\mu|,\lambda)\right\}\right) \label {distance1} \\
(p^z-H\lambda) d_z &= -2 r^2 \frac{m\gamma}{H}+2 (1-r^2)r\lambda\left\{\ln\left[\frac{m\gamma r-p^z+(1-r^2) H\lambda}{m\gamma_c r-p_c^z+(1-r^2) H\lambda}\right]-\bar B_z(r,|\mu|,\lambda)\right\}, \label{distance2}
\end{align}
Here, we have introduced the physical components relative separation between particle and anti-particle, $d_x = 2a(x-x_0)$, $d_z=2a(z-z_0)$, and we have used $A=m\gamma/H$. Inside of the logarithm, only the numerator is time dependent. In the denominator, the subindex in $\gamma_c$ and $p^z_c$ indicates that they are evaluated at the initial time when the pair is created. This is chosen so that the logarithm vanishes at that time. In going from Eqs. (\ref{pair1}-\ref{pair2}) to Eqs. (\ref{distance1}-\ref{distance2}) some constants have been absorbed into
the coefficients $\bar B_i$, which are still undetermined, and $r\lambda$ has been factored out in front of the curly brackets for later convenience.

The constants $\bar B_i$ can only depend on physical quantities at the time of nucleation. Therefore, we can express them in terms of the parameters $\lambda$ and $m$, as well as the kinematical properties $\gamma_c$ and $p^z_c$ at the time of nucleation. However, as we show in Appendix \ref{app:adiabatic}, $\gamma_c$ and $p^z_c$ are determined by $m,\lambda$ and $r$. Consequently we write $\bar B_i(r,|\mu|,\lambda)$, where we use the dimensionless combination 
$|\mu|=\sqrt{(m^2/H^2)+\lambda^2}$. To determine the coefficients $B_i$ we need additional input, as we now discuss.

First, we note that for $r=\pm 1$, we have $k_x=k_y=0$ and the dynamics is effectively 1+1 dimensional, with the charges moving along the direction of the electric field. In that case, there is an instanton solution in Euclidean de Sitter space, describing the nucleation of the pair. By analytic continuation, we can find the trajectories of the particle and antiparticle in the created pair, as well as their physical distance at any given moment of time. This was done in Ref. \cite{Frob:2014zka} 
(see Eq.~(A.1) in that reference), and the answer is given by
\begin{align}
(p^z-H\lambda) d_z = -2 \frac{m\gamma}{H}. \quad (r=\pm1)
\end{align}
This is in agreement with Eq.~\eqref{distance2}, provided that $B_z(r)$ does not have poles for $r^2=1$ . Also, for $\lambda=0$, there is no electric field and the system is rotationally invariant. Hence, the distance as a function of momentum should be the same in every direction. Squaring Eqs.~\eqref{distance1} and \eqref{distance2} 
and then adding, we have $p^2 d^2 = 4 m^2H^{-2} \gamma^2$ for all values of $r$, as expected.
These two observations justify our choice of including the unknown coefficients $\bar B_i$ inside of the curly brackets in Eqs.~\eqref{distance1} and \eqref{distance2}. Note that aside from $\lambda$, the parameter $r$ has also been factored out. This is simply to make the expression manifestly invariant under the simultaneous exchange of sign of $\lambda$ and $r$. This exchanges the role of particle and antiparticle and the the direction of motion relative to the electric field. 
It should be emphasized that in the two cases discussed here ($r=\pm1$ and/or $\lambda=0$) there are no ambiguities in the distance, since the relative position is fixed by analytic continuation of the Euclidean trajectory. 

Unfortunately, for $r\neq \pm1$ and $\lambda\neq 0$ there are no known instanton solutions. In fact, it seems unlikely that the instanton solutions exist for generic $r$ when $\lambda\neq 0$. The argument is the following. If we interpret the exponent of $|\beta_k|^2 \sim e^{-2\pi(|\mu|+\lambda r)}$, given in Eq.~\eqref{eq:bogo}, as the action of a would-be instanton, then we see that the only extrema as a function of $r$ are for $r=\pm 1$. For intermediate values of $r$, this action is linear in $r$ and would not be extremal with respect to small changes in the orientation of the orbit, which are parametrized by $r$. This is in contradiction with the assumption that there is an instanton solution, for which the action should be stationary. In the absence of instantons for generic $r$ and $\lambda$, we cannot determine the coefficients $\bar B_i$ semiclassically, at least not with much precision. 

Nonetheless, we can at least constrain its form, from the following arguments.
From Eq.~\eqref{distance2}, the separation in the $z$ direction at the time of pair creation is given by
\begin{align}
(p_c^z-H\lambda) d_{z,c} = -2 r^2 \frac{m\gamma_c}{H}-2 (1-r^2)r\lambda\bar B_z.
\end{align}
In the limit of a very large mass ($m/H\gg 1,\lambda$), we have $\gamma_c \sim 1$ and $p_c^z \sim mr$, while on physical grounds $d_z\lesssim r H^{-1}$. For that reason, it is clear that in this limit $B_z$ will be at most linear in $|\mu|$. Hence, assuming that the distance is analytic in all parameters, the leading behaviour in this limit is given by
\begin{align}
\bar B_z = |\mu| f_{z}(r) + O(\lambda r).
\end{align}
Here $f_{z}(r)$ is a smooth even function, with $|f_z|\lesssim 1$. A similar argument can be made for $\bar B_x$.
Hence, in the large mass $|\mu|\gg 1$ and weak field $\lambda\ll1$ limit, we have
\begin{align}
H \frac{k_c}{a} d_{z,c} \approx -2r m\gamma_c -2H (1-r^2)  |\mu| f_z(r) \lambda (1+ O(\lambda)). 
\end{align}
where $f_{z}(r)$ is a smooth even function, with $|f_z|\lesssim 1$. 

To conclude, we would like to explore the possibility of determining the distance between the particles in a pair based on analytic continuation of the classical trajectories. 
First, we note that 
\begin{equation}
(A+k\tau-r\lambda)(A-k\tau+r\lambda) = {m^2\over H^2}+ (1-r^2)\lambda^2 \equiv \alpha^2>0.
\end{equation}
%Note that $A$, given in (\ref{A2}) is a square root, and hence it is double valued. Now, we replace the logarithm in the semiclassical trajectory, which represents the solution for a positive value of $A$, by an analytic function that is real valued for both real branches of $A$ (that is, both for positive and negative $A$):
%\begin{equation}
%\ln\left[A+k \tau-r\lambda\right] \to {1\over 2} \lim_{\epsilon\to 0} \ln \left[(A+k\tau-r\lambda)^2 + \epsilon^2\right]=
%{1\over 2} \lim_{\epsilon\to 0} \ln \left[{A+k\tau-r\lambda\over A-k\tau-r\lambda}+ \epsilon^2\right]+ \ln \alpha.
%\end{equation}
%Then, the ratio between square brackets is always positive and non-vanishing, regardless of whether we take the positive or the negative branch
%for $A$, and the limit $\epsilon\to 0$ can be taken trivially in both cases. Consequently, dropping additive constants, 
Hence, the classical trajectory can be written as:
\begin{align}
k_z (z-z_0)
%\equiv\frac{k_z\,z-k_z\,\bar z}{2}
&= 
-r^2 A+(1-r^2)\frac{r\lambda}{2} \ln\left[\frac{A+k \tau-r\lambda}{A-k \tau+r\lambda}\right], \label{pair22}
\end{align}
and similarly for the $x$ coordinate. So far, we have assumed that $A$ is positive, in which case Eq. (\ref{pair22}) is the same as (\ref{pair2}), with $B_z=r\lambda\ln\alpha$. 
Note that $A^2 = \alpha^2 + (k\tau-r\lambda)^2>(k\tau-r\lambda)^2$, and therefore both numerator and denominator inside the square brackets in (\ref{pair22}) are positive. An interesting property of (\ref{pair22}), however, is that $(z-z_0)$ changes its overall sign under the replacement $A\to -A$, corresponding to the double valuedness of $A$ due to the square root in Eq. (\ref{A2}). In other words, Eq. (\ref{pair22}) is an analytic function which admits two real branches, representing the trajectories of a particle and that of an antiparticle. It is therefore very tempting to consider these to be the actual trajectories of the pair after nucleation. However, in this case the total semi-classical current \eqref{eq:totalsemi} would be positive definite (since $\bar f=0$). This would be in contradiction with the large mass limit result for the current Eq.~\eqref{eq:paircurrent} (once the EH terms have been subtracted). On the other hand, 
it should be noted that the two branches mentioned above cannot be obtained continuously from one another by analytic continuation, as is clear from (\ref{pair2}). When we go from positive $A$ to negative $A$ along a continuous path, the logarithm necessarily picks up an additive imaginary part. This difficulty is perhaps expected, since as we argued above there seem to be no instanton solutions for pairs with a transverse momentum.

%the embedding coordinates of de Sitter (in 5D Minkowski) that satisfy
%\begin{align}\label{eq:hyperboloid}
%\eta_{AB}X^AX^B=H^{-2}\qquad (A,B=0,...,4)
%\end{align}
%and $X^A$ are the coordinates in the 5D Minkowski space-time. They are related with the flat slicing by
%\begin{align}
%X^0&\equiv\frac{-1}{2\tau}\left(H^{-2}+\tau^2-R^2\right)\\
%X^4&\equiv\frac{-1}{2\tau}\left(H^{-2}-\tau^2+R^2\right)\\
%X^i&\equiv\frac{-x^i}{H\tau}\qquad (i=1,2,3)
%\end{align}
%where $R^2=x^2+y^2+z^2$. In the limit where $r=\pm1$ we see that the trajectories are given by
%\begin{align}
%\left(X^3\right)^2-\left(X^0\right)^2=R_0^2\equiv\frac{m^2}{m^2H^2+\lambda^2H^4}
%\end{align}
%which correspond to intersecting the hyperboloid \eqref{eq:hyperboloid} with the plane
%\begin{align}
%X^4=\mp w_0=\mp\left(H^{-2}-R_0^2\right)^{1/2}\,.
%\end{align}
%This case is the one schematically plotted in Fig.~\ref{fig:0}. When $\lambda=0$ we trivially recover the results of gravitational pair creation in de Sitter. This means that at least near these two limits we know that the particles will be nucleated at a distance
%\begin{align}
%d_z\equiv a\Delta z=\frac{2a}{k}\left\{\frac{-m\gamma r}{H}+\left(1-r^2\right)\lambda\ln\left[\frac{m\gamma }{H}-\frac{k}{aH}-\,r\lambda\right]\right\}\,.
%\end{align}


\begin{thebibliography}{100}

%\cite{Garriga:1994bm}
\bibitem{Garriga:1994bm}
  J.~Garriga,
 ``Pair production by an electric field in (1+1)-dimensional de Sitter space,''
  Phys.\ Rev.\ D {\bf 49} (1994) 6343.
  doi:10.1103/PhysRevD.49.6343
  %%CITATION = doi:10.1103/PhysRevD.49.6343;%%
  %62 citations counted in INSPIRE as of 13 Aug 2018


\bibitem{Frob:2014zka}
  M.~B.~Fröb, J.~Garriga, S.~Kanno, M.~Sasaki, J.~Soda, T.~Tanaka and A.~Vilenkin,
 ``Schwinger effect in de Sitter space,''
  JCAP {\bf 1404} (2014) 009
  %doi:10.1088/1475-7516/2014/04/009
  [arXiv:1401.4137 [hep-th]].
  %%CITATION = doi:10.1088/1475-7516/2014/04/009;%%
  %47 citations counted in INSPIRE as of 18 Jul 2018

  %\cite{Kobayashi:2014zza}
\bibitem{Kobayashi:2014zza}
  T.~Kobayashi and N.~Afshordi,
  ``Schwinger Effect in 4D de Sitter Space and Constraints on Magnetogenesis in the Early Universe,''
  JHEP {\bf 1410} (2014) 166
  %doi:10.1007/JHEP10(2014)166
  [arXiv:1408.4141 [hep-th]].
  %%CITATION = doi:10.1007/JHEP10(2014)166;%%
  %42 citations counted in INSPIRE as of 18 Jul 2018

  %\cite{Hayashinaka:2016qqn}
\bibitem{Hayashinaka:2016qqn}
  T.~Hayashinaka, T.~Fujita and J.~Yokoyama,
  ``Fermionic Schwinger effect and induced current in de Sitter space,''
  JCAP {\bf 1607} (2016) no.07,  010
  %doi:10.1088/1475-7516/2016/07/010
  [arXiv:1603.04165 [hep-th]].
  %%CITATION = doi:10.1088/1475-7516/2016/07/010;%%
  %20 citations counted in INSPIRE as of 18 Jul 2018
%\cite{Hayashinaka:2018amz}

%\cite{Hayashinaka:2016dnt}
\bibitem{Hayashinaka:2016dnt}
  T.~Hayashinaka and J.~Yokoyama,
 ``Point splitting renormalization of Schwinger induced current in de Sitter spacetime,''
  JCAP {\bf 1607} (2016) no.07,  012
%  doi:10.1088/1475-7516/2016/07/012
  [arXiv:1603.06172 [hep-th]].
  %%CITATION = doi:10.1088/1475-7516/2016/07/012;%%
  %14 citations counted in INSPIRE as of 26 Jul 2018


%\cite{Stahl:2015gaa}
\bibitem{Stahl:2015gaa}
  C.~Stahl, E.~Strobel and S.~S.~Xue,
  %``Fermionic current and Schwinger effect in de Sitter spacetime,''
  Phys.\ Rev.\ D {\bf 93} (2016) no.2,  025004
%  doi:10.1103/PhysRevD.93.025004
  [arXiv:1507.01686 [gr-qc]].
  %%CITATION = doi:10.1103/PhysRevD.93.025004;%%
  %24 citations counted in INSPIRE as of 01 Oct 2018

%\cite{Bavarsad:2016cxh}
\bibitem{Bavarsad:2016cxh}
  E.~Bavarsad, C.~Stahl and S.~S.~Xue,
  %``Scalar current of created pairs by Schwinger mechanism in de Sitter spacetime,''
  Phys.\ Rev.\ D {\bf 94} (2016) no.10,  104011
%  doi:10.1103/PhysRevD.94.104011
  [arXiv:1602.06556 [hep-th]].
  %%CITATION = doi:10.1103/PhysRevD.94.104011;%%
  %13 citations counted in INSPIRE as of 01 Oct 2018

%\cite{Bavarsad:2017oyv}
\bibitem{Bavarsad:2017oyv}
  E.~Bavarsad, S.~P.~Kim, C.~Stahl and S.~S.~Xue,
  %``Effect of a magnetic field on Schwinger mechanism in de Sitter spacetime,''
  Phys.\ Rev.\ D {\bf 97} (2018) no.2,  025017
%  doi:10.1103/PhysRevD.97.025017
  [arXiv:1707.03975 [hep-th]].
  %%CITATION = doi:10.1103/PhysRevD.97.025017;%%
  %10 citations counted in INSPIRE as of 01 Oct 2018

%\cite{Hayashinaka:2018amz}
\bibitem{Hayashinaka:2018amz}
  T.~Hayashinaka and S.~S.~Xue,
 ``Physical renormalization condition for de Sitter QED,''
  Phys.\ Rev.\ D {\bf 97} (2018) no.10,  105010
%  doi:10.1103/PhysRevD.97.105010
  [arXiv:1802.03686 [gr-qc]].
  %%CITATION = doi:10.1103/PhysRevD.97.105010;%%
  %3 citations counted in INSPIRE as of 26 Jul 2018

%\cite{Stahl:2018idd}
\bibitem{Stahl:2018idd}
  C.~Stahl,
  %``Schwinger effect impacting primordial magnetogenesis,''
  arXiv:1806.06692 [hep-th].
  %%CITATION = ARXIV:1806.06692;%%
  %2 citations counted in INSPIRE as of 01 Oct 2018

%%% "DETECTION"

%\cite{Neronov:1900zz}
\bibitem{Neronov:1900zz}
  A.~Neronov and I.~Vovk,
  ``Evidence for strong extragalactic magnetic fields from Fermi observations of TeV blazars,''
  Science {\bf 328} (2010) 73
  %doi:10.1126/science.1184192
  [arXiv:1006.3504 [astro-ph.HE]].
  %%CITATION = doi:10.1126/science.1184192;%%
  %482 citations counted in INSPIRE as of 26 Aug 2018
%\cite{Tavecchio:2010ja}
\bibitem{Tavecchio:2010ja}
  F.~Tavecchio, G.~Ghisellini, G.~Bonnoli and L.~Foschini,
  ``Extreme TeV blazars and the intergalactic magnetic field,''
  Mon.\ Not.\ Roy.\ Astron.\ Soc.\  {\bf 414} (2011) 3566
%  doi:10.1111/j.1365-2966.2011.18657.x
  [arXiv:1009.1048 [astro-ph.HE]].
  %%CITATION = doi:10.1111/j.1365-2966.2011.18657.x;%%
  %103 citations counted in INSPIRE as of 26 Aug 2018

%
\bibitem{Tavecchio:2010mk}
  F.~Tavecchio, G.~Ghisellini, L.~Foschini, G.~Bonnoli, G.~Ghirlanda and P.~Coppi,
  ``The intergalactic magnetic field constrained by Fermi/LAT observations of the TeV blazar 1ES 0229+200,''
  Mon.\ Not.\ Roy.\ Astron.\ Soc.\  {\bf 406} (2010) L70
%  doi:10.1111/j.1745-3933.2010.00884.x
  [arXiv:1004.1329 [astro-ph.CO]].
  %%CITATION = doi:10.1111/j.1745-3933.2010.00884.x;%%
  %225 citations counted in INSPIRE as of 26 Aug 2018

%\cite{Taylor:2011bn}
\bibitem{Taylor:2011bn}
  A.~M.~Taylor, I.~Vovk and A.~Neronov,
  ``Extragalactic magnetic fields constraints from simultaneous GeV-TeV observations of blazars,''
  Astron.\ Astrophys.\  {\bf 529} (2011) A144
%  doi:10.1051/0004-6361/201116441
  [arXiv:1101.0932 [astro-ph.HE]].
  %%CITATION = doi:10.1051/0004-6361/201116441;%%
  %168 citations counted in INSPIRE as of 26 Aug 2018

%%% REVIEWS


  %\cite{Kandus:2010nw}
\bibitem{Kandus:2010nw} 
  A.~Kandus, K.~E.~Kunze and C.~G.~Tsagas,
  ``Primordial magnetogenesis,''
  Phys.\ Rept.\  {\bf 505}, 1 (2011)
  %doi:10.1016/j.physrep.2011.03.001
  [arXiv:1007.3891 [astro-ph.CO]].
  %%CITATION = doi:10.1016/j.physrep.2011.03.001;%%
  %144 citations counted in INSPIRE as of 18 Nov 2015
  
  %\cite{Durrer:2013pga}
\bibitem{Durrer:2013pga} 
  R.~Durrer and A.~Neronov,
  ``Cosmological Magnetic Fields: Their Generation, Evolution and Observation,''
  Astron.\ Astrophys.\ Rev.\  {\bf 21}, 62 (2013)
 % doi:10.1007/s00159-013-0062-7
  [arXiv:1303.7121 [astro-ph.CO]].
  %%CITATION = doi:10.1007/s00159-013-0062-7;%%
  %110 citations counted in INSPIRE as of 18 Nov 2015
  
%\cite{Subramanian:2015lua}
\bibitem{Subramanian:2015lua}
  K.~Subramanian,
  ``The origin, evolution and signatures of primordial magnetic fields,''
  Rept.\ Prog.\ Phys.\  {\bf 79} (2016) no.7,  076901
%  doi:10.1088/0034-4885/79/7/076901
  [arXiv:1504.02311 [astro-ph.CO]].
  %%CITATION = doi:10.1088/0034-4885/79/7/076901;%%
  %86 citations counted in INSPIRE as of 26 Aug 2018


%%% NOTEWORTHY PAPERS

%\cite{Turner:1987bw}
\bibitem{Turner:1987bw}
  M.~S.~Turner and L.~M.~Widrow,
  ``Inflation Produced, Large Scale Magnetic Fields,''
  Phys.\ Rev.\ D {\bf 37} (1988) 2743.
%  doi:10.1103/PhysRevD.37.2743
  %%CITATION = doi:10.1103/PhysRevD.37.2743;%%
  %669 citations counted in INSPIRE as of 26 Aug 2018


%\cite{Ratra:1991bn}\cite{Turner:1987bw,Ratra:1991bn,Bekenstein:1982eu,Bamba:2006ga,Demozzi:2009fu}
\bibitem{Ratra:1991bn}
  B.~Ratra,
  ``Cosmological 'seed' magnetic field from inflation,''
  Astrophys.\ J.\  {\bf 391} (1992) L1.
%  doi:10.1086/186384
  %%CITATION = doi:10.1086/186384;%%
  %570 citations counted in INSPIRE as of 26 Aug 2018

%\cite{Bekenstein:1982eu}
\bibitem{Bekenstein:1982eu}
  J.~D.~Bekenstein,
  ``Fine Structure Constant: Is It Really a Constant?,''
  Phys.\ Rev.\ D {\bf 25} (1982) 1527.
%  doi:10.1103/PhysRevD.25.1527
  %%CITATION = doi:10.1103/PhysRevD.25.1527;%%
  %316 citations counted in INSPIRE as of 26 Aug 2018

%\cite{Bamba:2006ga}
\bibitem{Bamba:2006ga} 
  K.~Bamba and M.~Sasaki,
  ``Large-scale magnetic fields in the inflationary universe,''
  JCAP {\bf 0702}, 030 (2007)
%  doi:10.1088/1475-7516/2007/02/030
%  [astro-ph/0611701].
  %%CITATION = doi:10.1088/1475-7516/2007/02/030;%%

%\cite{Demozzi:2009fu}
\bibitem{Demozzi:2009fu} 
  V.~Demozzi, V.~Mukhanov and H.~Rubinstein,
  ``Magnetic fields from inflation?,''
  JCAP {\bf 0908}, 025 (2009)
  %doi:10.1088/1475-7516/2009/08/025
  [arXiv:0907.1030 [astro-ph.CO]].
  %%CITATION = doi:10.1088/1475-7516/2009/08/025;%%

  
  %\cite{Suyama:2012wh}
\bibitem{Suyama:2012wh} 
  T.~Suyama and J.~Yokoyama,
  ``Metric perturbation from inflationary magnetic field and generic bound on inflation models,''
  Phys.\ Rev.\ D {\bf 86}, 023512 (2012)
%  doi:10.1103/PhysRevD.86.023512
  [arXiv:1204.3976 [astro-ph.CO]].
  %%CITATION = doi:10.1103/PhysRevD.86.023512;%%
  %23 citations counted in INSPIRE as of 26 Nov 2015
  
  %\cite{Fujita:2012rb}
\bibitem{Fujita:2012rb} 
  T.~Fujita and S.~Mukohyama,
  ``Universal upper limit on inflation energy scale from cosmic magnetic field,''
  JCAP {\bf 1210}, 034 (2012)
  %doi:10.1088/1475-7516/2012/10/034
  [arXiv:1205.5031 [astro-ph.CO]].
  %%CITATION = doi:10.1088/1475-7516/2012/10/034;%%
  %46 citations counted in INSPIRE as of 18 Nov 2015
  
  %\cite{Fujita:2013pgp}
\bibitem{Fujita:2013pgp} 
  T.~Fujita and S.~Yokoyama,
  ``Higher order statistics of curvature perturbations in IFF model and its Planck constraints,''
  JCAP {\bf 1309}, 009 (2013)
 % doi:10.1088/1475-7516/2013/09/009
  [arXiv:1306.2992 [astro-ph.CO]].
  %%CITATION = doi:10.1088/1475-7516/2013/09/009;%%
  %26 citations counted in INSPIRE as of 18 Nov 2015
  
  %\cite{Fujita:2014sna}
\bibitem{Fujita:2014sna} 
  T.~Fujita and S.~Yokoyama,
  ``Critical constraint on inflationary magnetogenesis,''
  JCAP {\bf 1403}, 013 (2014)
  [JCAP {\bf 1405}, E02 (2014)]
  %doi:10.1088/1475-7516/2014/03/013, 10.1088/1475-7516/2014/05/E02
  [arXiv:1402.0596 [astro-ph.CO]].
  %%CITATION = doi:10.1088/1475-7516/2014/03/013, 10.1088/1475-7516/2014/05/E02;%%
  %17 citations counted in INSPIRE as of 18 Nov 2015
  
%\cite{Ferreira:2013sqa}
  \bibitem{Ferreira:2013sqa}
  R.~J.~Z.~Ferreira, R.~K.~Jain and M.~S.~Sloth,
  ``Inflationary magnetogenesis  without the strong coupling problem,''
  JCAP {\bf 1310} (2013) 004
 % doi:10.1088/1475-7516/2013/10/004
  [arXiv:1305.7151 [astro-ph.CO]].
  %%CITATION = doi:10.1088/1475-7516/2013/10/004;%%
  %46 citations counted in INSPIRE as of 25 May 2016
  
  %\cite{Ferreira:2014hma}
\bibitem{Ferreira:2014hma} 
  R.~J.~Z.~Ferreira, R.~K.~Jain and M.~S.~Sloth,
  ``Inflationary Magnetogenesis without the Strong Coupling Problem II: Constraints from CMB anisotropies and B-modes,''
  JCAP {\bf 1406}, 053 (2014)
 % doi:10.1088/1475-7516/2014/06/053
  [arXiv:1403.5516 [astro-ph.CO]].
  %%CITATION = doi:10.1088/1475-7516/2014/06/053;%%
  %18 citations counted in INSPIRE as of 18 Nov 2015
  
%\cite{Green:2015fss}
\bibitem{Green:2015fss}
  D.~Green and T.~Kobayashi,
  ``Constraints on Primordial Magnetic Fields from Inflation,''
  JCAP {\bf 1603} (2016) no.03,  010
 % doi:10.1088/1475-7516/2016/03/010
  [arXiv:1511.08793 [astro-ph.CO]].
  %%CITATION = doi:10.1088/1475-7516/2016/03/010;%%
  %15 citations counted in INSPIRE as of 26 Aug 2018

%\cite{Fujita:2015iga}
\bibitem{Fujita:2015iga}
  T.~Fujita, R.~Namba, Y.~Tada, N.~Takeda and H.~Tashiro,
  ``Consistent generation of magnetic fields in axion inflation models,''
  JCAP {\bf 1505} (2015) no.05,  054
%  doi:10.1088/1475-7516/2015/05/054
  [arXiv:1503.05802 [astro-ph.CO]].
  %%CITATION = doi:10.1088/1475-7516/2015/05/054;%%
  %39 citations counted in INSPIRE as of 26 Aug 2018

%\cite{Domenech:2015zzi}
\bibitem{Domenech:2015zzi}
  G.~Dom{\`e}nech, C.~Lin and M.~Sasaki,
  ``Inflationary Magnetogenesis with Broken Local U(1) Symmetry,''
  EPL {\bf 115} (2016) no.1,  19001
%  doi:10.1209/0295-5075/115/19001
  [arXiv:1512.01108 [astro-ph.CO]].
  %%CITATION = doi:10.1209/0295-5075/115/19001;%%
  %10 citations counted in INSPIRE as of 26 Aug 2018

%\cite{Fujita:2016qab}
\bibitem{Fujita:2016qab}
  T.~Fujita and R.~Namba,
  ``Pre-reheating Magnetogenesis in the Kinetic Coupling Model,''
  Phys.\ Rev.\ D {\bf 94} (2016) no.4,  043523
%  doi:10.1103/PhysRevD.94.043523
  [arXiv:1602.05673 [astro-ph.CO]].
  %%CITATION = doi:10.1103/PhysRevD.94.043523;%%
  %17 citations counted in INSPIRE as of 26 Aug 2018

%\cite{Sharma:2017eps}
\bibitem{Sharma:2017eps}
  R.~Sharma, S.~Jagannathan, T.~R.~Seshadri and K.~Subramanian,
  ``Challenges in Inflationary Magnetogenesis: Constraints from Strong Coupling, Backreaction and the Schwinger Effect,''
  Phys.\ Rev.\ D {\bf 96} (2017) no.8,  083511
%  doi:10.1103/PhysRevD.96.083511
  [arXiv:1708.08119 [astro-ph.CO]].
  %%CITATION = doi:10.1103/PhysRevD.96.083511;%%
  %8 citations counted in INSPIRE as of 26 Aug 2018

%\cite{Motta:2012rn}
\bibitem{Motta:2012rn}
L.~Motta and R.~R.~Caldwell,
``Non-Gaussian features of primordial magnetic fields in power-law inflation,''
Phys.\ Rev.\ D {\bf 85} (2012) 103532
%doi:10.1103/PhysRevD.85.103532
[arXiv:1203.1033 [astro-ph.CO]].
%%CITATION = doi:10.1103/PhysRevD.85.103532;%%
%27 citations counted in INSPIRE as of 26 May 2016

\bibitem{Giovannini:2013rza} 
  M.~Giovannini,
  ``Inflationary susceptibilities, duality, and large-scale magnetic field generation,''
  Phys.\ Rev.\ D {\bf 88}, no. 8, 083533 (2013)
  %doi:10.1103/PhysRevD.88.083533
  [arXiv:1310.1802 [hep-th]].
  %%CITATION = doi:10.1103/PhysRevD.88.083533;%%
  %9 citations counted in INSPIRE as of 27 May 2016

\bibitem{Tasinato:2014fia}
 G.~Tasinato,
 ``A scenario for inflationary magnetogenesis without strong coupling problem,''
 JCAP {\bf 1503} (2015) 040
 %doi:10.1088/1475-7516/2015/03/040
 [arXiv:1411.2803 [hep-th]].
 %%CITATION = doi:10.1088/1475-7516/2015/03/040;%%
 

%\cite{Lozanov:2018kpk}\cite{Ito:2017bnn}
\bibitem{Lozanov:2018kpk}
  K.~D.~Lozanov, A.~Maleknejad and E.~Komatsu,
 ``Schwinger Effect by an $SU(2)$ Gauge Field during Inflation,''
  arXiv:1805.09318 [hep-th].
  %%CITATION = ARXIV:1805.09318;%%
  %4 citations counted in INSPIRE as of 13 Aug 2018

%\cite{Watanabe:2009ct}
\bibitem{Watanabe:2009ct}
  M.~a.~Watanabe, S.~Kanno and J.~Soda,
  ``Inflationary Universe with Anisotropic Hair,''
  Phys.\ Rev.\ Lett.\  {\bf 102} (2009) 191302
  doi:10.1103/PhysRevLett.102.191302
  [arXiv:0902.2833 [hep-th]].
  %%CITATION = doi:10.1103/PhysRevLett.102.191302;%%
  %249 citations counted in INSPIRE as of 01 Aug 2018

%\cite{Ito:2017bnn}
\bibitem{Ito:2017bnn}
  A.~Ito and J.~Soda,
  ``Anisotropic Constant-roll Inflation,''
  Eur.\ Phys.\ J.\ C {\bf 78} (2018) no.1,  55
  doi:10.1140/epjc/s10052-018-5534-5
  [arXiv:1710.09701 [hep-th]].
  %%CITATION = doi:10.1140/epjc/s10052-018-5534-5;%%
  %13 citations counted in INSPIRE as of 13 Aug 2018


%\cite{Kitamoto:2018htg}
\bibitem{Kitamoto:2018htg}
  H.~Kitamoto,
  ``Schwinger Effect in Inflaton-Driven Electric Field,''
  arXiv:1807.03753 [hep-th].
  %%CITATION = ARXIV:1807.03753;%%


%\cite{Stahl:2016geq}
\bibitem{Stahl:2016geq}
  C.~Stahl and S.~S.~Xue,
  %``Schwinger effect and backreaction in de Sitter spacetime,''
  Phys.\ Lett.\ B {\bf 760} (2016) 288
%  doi:10.1016/j.physletb.2016.07.011
  [arXiv:1603.07166 [hep-th]].
  %%CITATION = doi:10.1016/j.physletb.2016.07.011;%%
  %10 citations counted in INSPIRE as of 01 Oct 2018


%\cite{Sobol:2018djj}
\bibitem{Sobol:2018djj}
  O.~O.~Sobol, E.~V.~Gorbar, M.~Kamarpour and S.~I.~Vilchinskii,
  %``Influence of back-reaction of electric fields and Schwinger effect on inflationary magnetogenesis,''
  arXiv:1807.09851 [hep-ph].
  %%CITATION = ARXIV:1807.09851;%%



\bibitem{birefringence} 
  J.~S.~Heyl and L.~Hernquist,
  ``Birefringence and dichroism of the QED vacuum,''
  J.\ Phys.\ A {\bf 30}, 6485 (1997)
  %doi:10.1088/0305-4470/30/18/022
  [hep-ph/9705367].
  %%CITATION = doi:10.1088/0305-4470/30/18/022;%%

%\cite{Cai:2014qba}
\bibitem{Cai:2014qba}
  R.~G.~Cai and S.~P.~Kim,
  %``One-Loop Effective Action and Schwinger Effect in (Anti-) de Sitter Space,''
  JHEP {\bf 1409} (2014) 072
  doi:10.1007/JHEP09(2014)072
  [arXiv:1407.4569 [hep-th]].
  %%CITATION = doi:10.1007/JHEP09(2014)072;%%
  %25 citations counted in INSPIRE as of 01 Oct 2018


%\cite{Dunne:2004nc}
\bibitem{Dunne:2004nc}
  G.~V.~Dunne,
  ``Heisenberg-Euler effective Lagrangians: Basics and extensions,''
  In Shifman, M. (ed.) et al.: From fields to strings, vol. 1 445-522
 % doi:10.1142/9789812775344_0014
  [hep-th/0406216].
  %%CITATION = doi:10.1142/9789812775344_0014;%%
  %246 citations counted in INSPIRE as of 18 Jul 2018

%\cite{Codello:2012kq}
\bibitem{Codello:2012kq}
  A.~Codello and O.~Zanusso,
  ``On the non-local heat kernel expansion,''
  J.\ Math.\ Phys.\  {\bf 54} (2013) 013513
  doi:10.1063/1.4776234
  [arXiv:1203.2034 [math-ph]].
  %%CITATION = doi:10.1063/1.4776234;%%
  %25 citations counted in INSPIRE as of 13 Aug 2018

%\cite{Barvinsky:1990up}
\bibitem{Barvinsky:1990up}
  A.~O.~Barvinsky and G.~A.~Vilkovisky,
  ``Covariant perturbation theory. 2: Second order in the curvature. General algorithms,''
  Nucl.\ Phys.\ B {\bf 333} (1990) 471.
  doi:10.1016/0550-3213(90)90047-H
  %%CITATION = doi:10.1016/0550-3213(90)90047-H;%%
  %225 citations counted in INSPIRE as of 13 Aug 2018


%\cite{Srednicki:2007qs}
\bibitem{Srednicki:2007qs} 
  M.~Srednicki,
  ``Quantum field theory,'' Cambridge University Press, 2007.
  %%CITATION = INSPIRE-752478;%%
  %268 citations counted in INSPIRE as of 22 Aug 2018


%\cite{Bastianelli:2008cu}
\bibitem{Bastianelli:2008cu}
  F.~Bastianelli, J.~M.~Davila and C.~Schubert,
 ``Gravitational corrections to the Euler-Heisenberg Lagrangian,''
  JHEP {\bf 0903} (2009) 086
 % doi:10.1088/1126-6708/2009/03/086
  [arXiv:0812.4849 [hep-th]].
  %%CITATION = doi:10.1088/1126-6708/2009/03/086;%%
  %24 citations counted in INSPIRE as of 18 Jul 2018


%\cite{Garriga:2012pk}
\bibitem{Garriga:2012pk}
  J.~Garriga and A.~Vilenkin,
  %``Living with ghosts in Lorentz invariant theories,''
  JCAP {\bf 1301} (2013) 036
  doi:10.1088/1475-7516/2013/01/036
  [arXiv:1202.1239 [hep-th]].
  %%CITATION = doi:10.1088/1475-7516/2013/01/036;%%
  %21 citations counted in INSPIRE as of 13 Sep 2018


  %\cite{Bicak:2005yt}
\bibitem{Bicak:2005yt}
  J.~Bicak and P.~Krtous,
  ``Fields of accelerated sources: Born in de Sitter,''
  J.\ Math.\ Phys.\  {\bf 46} (2005) 102504
  doi:10.1063/1.2009647
  [gr-qc/0602009].
  %%CITATION = doi:10.1063/1.2009647;%%
  %19 citations counted in INSPIRE as of 13 Aug 2018


%%%%%%%%%%%%%%%%%%%%%%%%%%%%%%%%%%%%%%%%%%%%%%%%%%%%%%%%%%%%
%%%%%%%%%%%%%%%%%%%%%%%%%%%%%%%%%%%%%%%%%%%%%%%%%%%%%%%%%%%%
%%%%%%%%%%%%%%%%%%%%%%%%%%%%%%%%%%%%%%%%%%%%%%%%%%%%%%%%%%%%
\end{thebibliography}
\end{document}